\documentclass[aps,prl,reprint,superscriptaddress]{revtex4-2}
\usepackage[colorlinks,linkcolor=blue,anchorcolor=blue,citecolor=blue]{hyperref}

\usepackage{graphicx,dcolumn,bm,booktabs, array, float, tabularx, booktabs, lipsum, amsmath,multirow,siunitx, xcolor,lineno,tikz,soul,autobreak}
\usepackage[utf8]{inputenc}
\usepackage[T1]{fontenc}
\usepackage[version=4]{mhchem}
\usepackage{color}
\usepackage{overpic}

\newcommand{\ee}{\mathrm{e}}
\newcommand{\mean}[1]{\left\langle  #1 \right\rangle }

\usepackage{titlesec}
\titleformat{\section} 
  [runin] 
  {\normalfont\itshape} 
  {} 
  {0pt} 
  {} 
\titlespacing*{\section} 
  {1em} 
  {0pt} 
  {0em} 
\begin{document}
\title{Multi-range fractional model for convective atmospheric surface-layer turbulence}

\author{Fei-Chi Zhang}
\affiliation{Department of Mechanics and Engineering Science at College of Engineering, and State Key Laboratory for Turbulence and Complex Systems, Peking University, Beijing 100871, P. R. China}
\author{Jin-Han Xie}
\email{jinhanxie@pku.edu.cn}
\affiliation{Department of Mechanics and Engineering Science at College of Engineering, and State Key Laboratory for Turbulence and Complex Systems, Peking University, Beijing 100871, P. R. China}
\author{Xiaojing Zheng}
\email{xjzheng@xidian.edu.cn}
\affiliation{Research Center for Applied Mechanics, Xidian University, Xi’an 710071, P. R. China}

\date{\today}

\begin{abstract}
We develop a multi-range fractional (MRF) model to capture the turbulent spectrum consisting of multiple self-similar ranges impacted by multiple effects.
The MRF model is validated using long-term observational atmospheric surface layer data from Qingtu lake with extreme Reynolds numbers up to Re$_\tau\sim O(10^6)$.
The spectral exponent in each range and the transition scales between different ranges are solo parameters in the MRF model and are identified for streamwise velocity, vertical velocity, and temperature, and they update the quantifications in the multi-point Monin-Obukhov theory.
Therefore, based on the MRF model and considering the consistency between the turbulent spectrum and variance, we propose an expression for the vertical dependence of the streamwise velocity variance that is inadequately described by the Monin-Obukhov similarity theory. 
The MRF model provides a new method to analyze and quantify turbulent data, and as a time-series model, it enables the generation of synthetic turbulent data.

\end{abstract}

\maketitle
\section{Introduction. }---
There are two types of complexities in turbulence: multiscale, which stems from the nonlinear nature of fluid motion, and multi-effects due to different environments. 
The multiscale behaviour can be captured by scalings since \citet{kolmogorov1941local}.
And we can capture these scaling behaviours using fractional Brownian motion \cite{Levy1965,mandelbrot1968fractional, friedrich2020stochastic}.
Multi-effects manifest themselves by different regions and ranges have different scaling behaviours, whose exponents and transition depend on factors of environments.
This letter focuses on turbulence in the atmospheric surface layer (ASL), the lowest part of the troposphere, which plays a critical role in modelling near-surface turbulence involving complex interactions between thermal stratification and wind shear \cite{Hoegstroem1996}. 
ASL turbulence influences numerous environmental and meteorological processes, including numerical weather prediction \cite{munoz2014bridging}, climate modelling \cite{mcgrath2015impact, salesky2020coherent}, and wind energy system design \cite{wachter2012turbulent}.

To capture the competing effects of shear and buoyancy, Monin-Obukhov similarity theory (MOST) \cite{monin1954basic} has been widely used under various stability conditions \cite{foken200650}.
In this theory, the thermal stability parameter $z/L$, where $z$ is the distance to the ground and $L$ is the Obukhov length measures the relative strength between shear production versus buoyancy effects, and the statistical mean profiles of wind speed, temperature, and turbulence intensity are proposed to be functions of $z/L$ only \cite{foken200650, wyngaard2010turbulence}.
Mean profiles in the shear-dominate lower region are harder to express analytically.
Based on statistical symmetry explored by Lie group analysis, the scaling of statistical quantities, including mean flow and moments, can be locally expressed \cite{Oberlack2022}.
\citet{She2010} push the theoretical development to capture multilayer structures and particularly an accurate capture of transitions between different layers.

The dynamics of the ASL are inherently multiscale, ranging from very-large-scale motions \cite{balakumar2007large,wang2016very} to the smallest viscous scales. 
Understanding these multiscale behaviours is crucial for accurately characterizing and modelling the turbulent processes in the ASL, as they influence the distribution and mixing of turbulent energy. 
The complex interaction between multiscale motions results in scaling properties of spectra and structure functions. 
At small scales, where the wall and buoyancy effects are subdominant, Kolmogorov scaling, $k_x^{-5/3}$ with $k_x$ the streamwise wavenumber exits \cite{de2015scaling}.
In shear-dominated ASL, a $k^{-1}_x$ scaling is observed in the energy spectrum at low streamwise wavenumbers $k_x$ \cite{Nickels2005}, corresponding to a $\ln{r}$ behaviour with $r$ the distance between two measured points for the structure functions \cite{davidson2006logarithmic, de2015scaling}.When buoyancy becomes significant, the spectral scaling shifts to $k^{-5/3}_x$ \cite{yaglom1994fluctuation, Kader1989}, which associates with an $r^{2/3}$ scaling for the second-order structure function in physical space \cite{chamecki2017scaling}. 
To capture the transitions between neutral and convective ASL spectra, \citet{tong2015multipoint} proposed the multi-point Monin-Obukhov (MMO) theory, which introduces three distinct power-law scalings with specific scale ranges and power exponents.

To fully capture the dynamics of ASL turbulence, it is necessary to consider variations in both vertical and streamwise directions, along with the multiscale effects. 
There are main problems to be solved: (i) How to determine the scaling exponents from data? (ii) How to analytically capture the transition between different ranges?
In this work, we propose a new statistical model that integrates these aspects: the multi-range fractional integrated (MRF) model. 
This model bases on two foundations: (i) Following the statistical understanding of non-equilibrium open systems, there exists a finite number of statistical states that form a multi-range picture, with each range corresponding to certain characteristic physical processes. 
(ii) Each range is characterized by a set of self-similar structures, quantified by its fractal dimension, and is described by fractional Brownian motion. 
Applying the MRF model to convective ASL provides a framework for ASL turbulence by incorporating the strengths of both MOST and MMO, while also addressing their respective limitations, such as disconnect between the one-point statistics of MOST and the multi-point framework of MMO, and the failure of MOST for streamwise velocity variances. 
In addition, as a time series stochastic model \cite{friedrich2020stochastic}, the MRF model enables the analysis and quantification of complex turbulent data.

\section{Multi-range fractional model. }---
The long-range memory of turbulent motions can be characterized by the Hurst exponent, which corresponds to a fractal dimension \cite{Voss1986}. 
However, the Hurst exponent only accounts for long-range correlations of a single self-similar motion. 
In the ASL, turbulent motions occur across multiple characteristic scales, such as attached eddies and very-large-scale motions, each with distinct correlation properties. 
As a result, a single Hurst exponent or fractal dimension alone is insufficient to describe the multi-range nature of these motions. 
To capture multi-range effects, we propose the multi-range fractional (MRF) model to characterize the scale-dependent fractal behaviour:
\begin{equation}\label{MRF}
	\prod \limits_{i=1}^{N}
	 (1-\ee^{-\lambda_i} \mathcal{B})^{d_i-d_{i-1}} 
	 u_t =\epsilon_t, 
\end{equation}
where $N$ represents the number of characteristic scales, $\mathcal{B}$ is the lag operator s.t. $\mathcal{B}u_t=u_{t-1}$, $\lambda_i>0$ represent the characteristic scales, $d_i$ are the fractional orders of differentiation, and $\epsilon_t$ is an uncorrelated random variable, which is assumed to be white noise for simplicity. 
The MRF model is stationary, casual, and invertible, which is a generalization of the tempered fractional integration model \cite{meerschaert2014tempered}. 
Details of MRF model are shown in Supplemental Material Part A. 
\begin{figure}
\includegraphics[width=\linewidth]{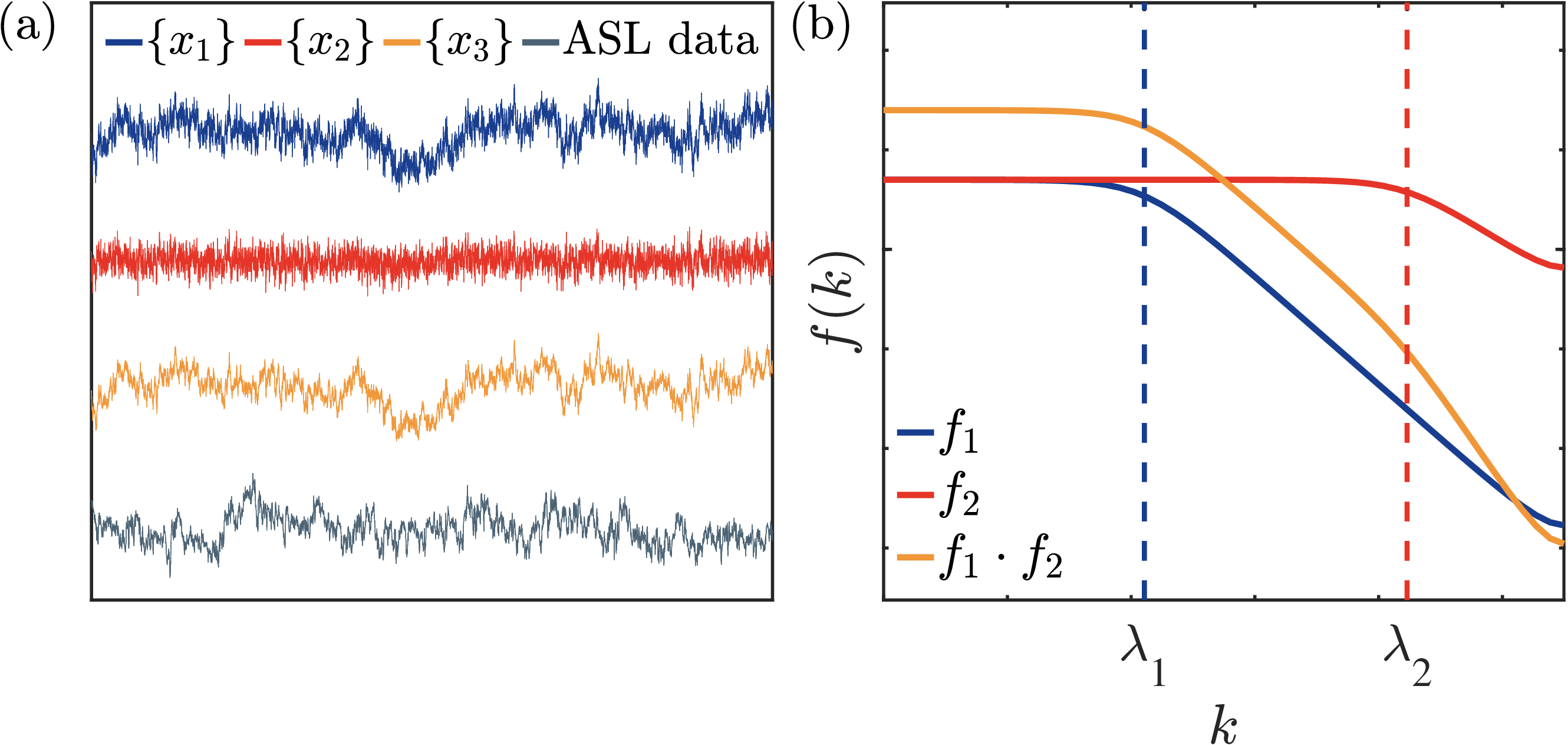}%
\caption{
Time series and spectra of the MRF model for $N=1$ and $N=2$. 
Time series $\{x_1\}$, $\{x_2\}$, and $\{x_3\}$ in (a) correspond to the model spectra $f_1$, $f_2$, and $f_1\cdot f_2$ in (b), respectively. 
Integrating the MRF model from $N=1$ to $N=2$ captures the multiscale and multi-range behaviour effectively. 
The synthetic time series appears similar to the ASL data.}
\label{fig_MRF}
\end{figure}

A great property of the MRF model is its analytical expression for energy spectrum:
\begin{equation}\label{MRF_spectrum}
	f(k)= \frac{\sigma_\epsilon^2}{2\pi} 
	\prod_{i=1}^{N}\left(1-2\ee^{-\lambda_i}\cos k+\ee^{-2\lambda_i}\right)^{-(d_i-d_{i-1})},
\end{equation}
where $\sigma_\epsilon^2$ is the variance of $\epsilon_t$, and $k \in [-\pi, \pi]$ represents the nondimensional frequency.
When $k<\min\{ \lambda_i\}$, $f$ approaches a constant value of $\sigma_\epsilon^2 \prod_{i=1}^{N}\left(1-\ee^{-\lambda_i}\right)^{-2(d_i-d_{i-1})}/(2\pi)$, and when $\lambda_M<k<\lambda_{M+1}$ with $M = 1, 2, \dots , N-1$, asymptotically we obtain
\begin{equation}\label{asympotic}
    \begin{aligned}
    f(k) &\approx \frac{\sigma_\epsilon^2}{2\pi}\prod_{j=M+1}^{N} \lambda_j^{-2(d_j-d_{j-1})}
    \prod_{i=1}^{M} (k^2+\lambda_i^2)^{-(d_i-d_{i-1})} \notag\\
    &\sim k^{-2d_M}. 
\end{aligned}
\end{equation}
Thus, the MRF captures the multi-range scalings with exponents $-2d_i$ within ranges divided by $\lambda_i$.
Noting that expressions with similar asymptotic behaviours have been applied in turbulence research empirically \cite{Batchelor1951Pressure,pope2000turbulent} or based on symmetry arguments \cite{She2010}, in comparison, our expression (\ref{MRF_spectrum}) is analytically obtained from a stochastic time series model.

The MRF model integrates stochastic processes associated with different characteristic scales and scaling laws. 
In Fig. \ref{fig_MRF}(a), ${x_1}$ and ${x_2}$ represent a single-range time series with one spectral exponent, which can be captured by tempered fractional Brownian motion \cite{meerschaert2013tempered}. 
Using the MRF model, we can combine ${x_1}$ and ${x_2}$ to obtain a time series ${x_3}$ with scale-dependent correlations: at large scales, ${x_3}$ exhibits the same correlation as ${x_1}$, while at small scales, the correlation of ${x_3}$ is influenced by both ${x_1}$ and ${x_2}$. 
The resulting synthetic time series ${x_3}$ well capture the key features of the ASL data.
Fig. \ref{fig_MRF}(b) shows spectra of time series composed of two single-range fractional operators.
The transition scales $\lambda_i$ of single-range models remain in the composed spectrum, and the scaling exponents of the latter follow (\ref{asympotic}).
Therefore, the model parameters $\lambda_i$ and $d_i$ capture characteristic scales and spectrum exponents. 

\begin{figure}[htbp]
  \centering
  \begin{overpic}[width=0.238\textwidth]{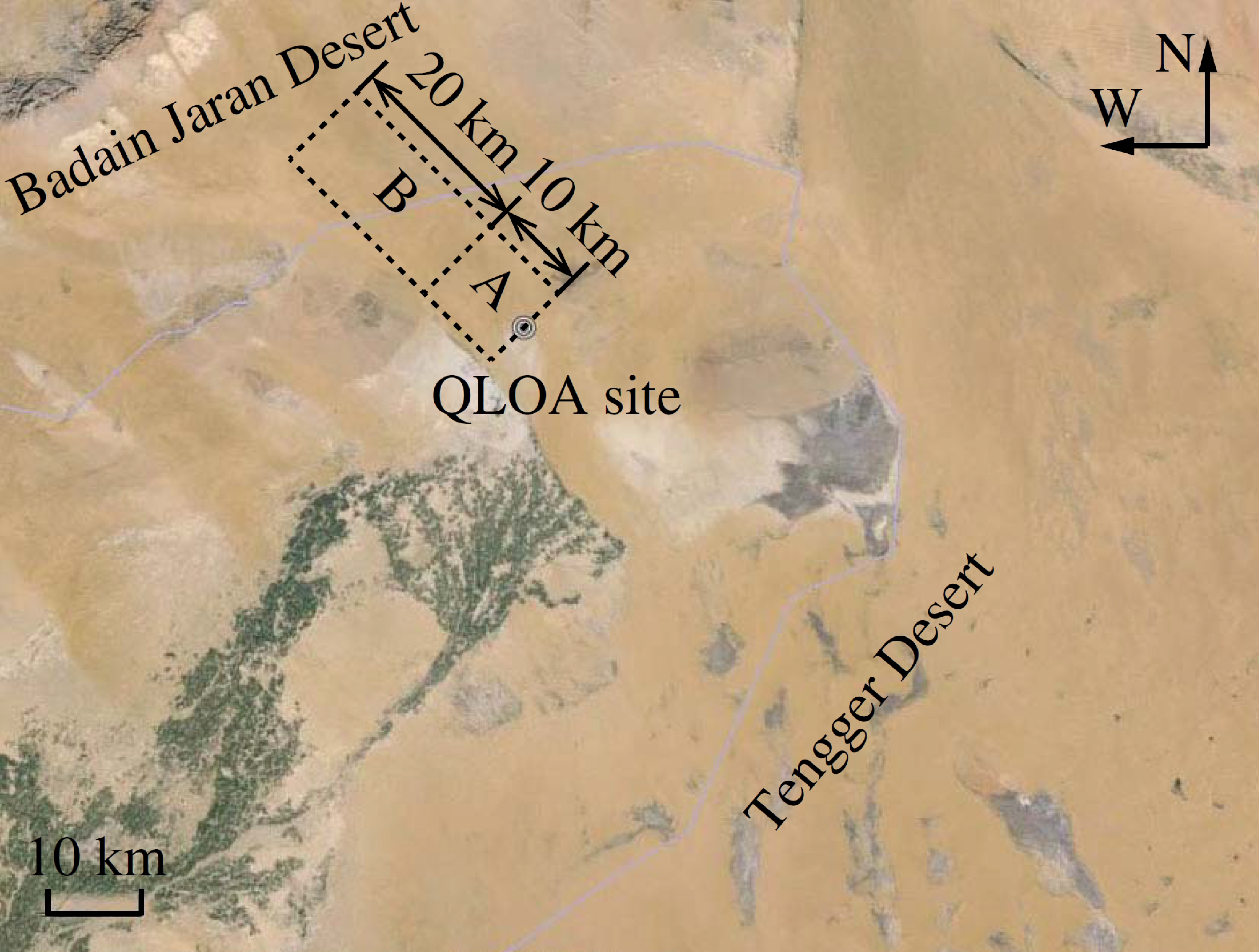}
    \put(1,68){(a)}
  \end{overpic}
  \begin{overpic}[width=0.238\textwidth]{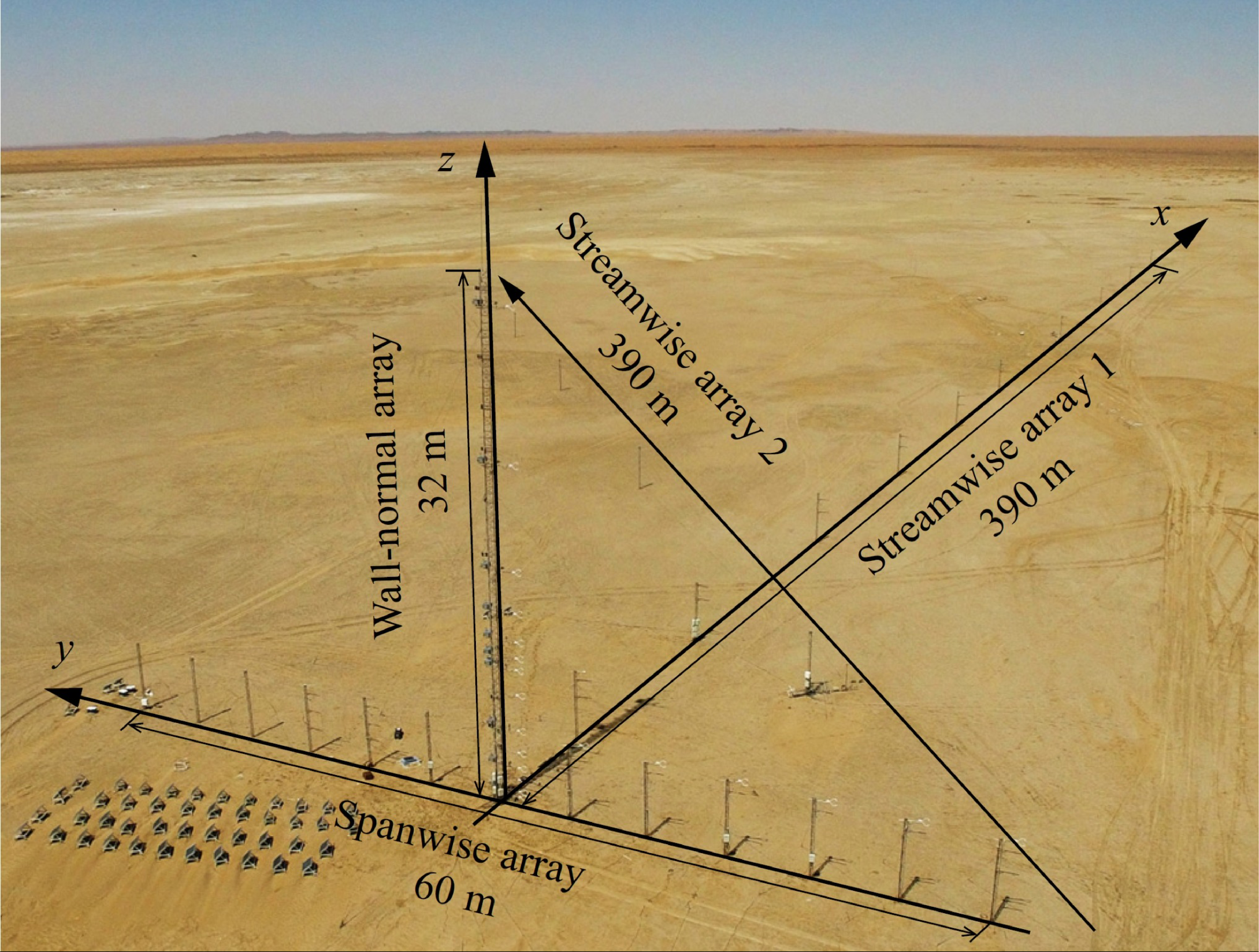}
    \put(1,68){(b)}
  \end{overpic}
  \caption{(a) Satellite photograph and (b) panoramic view of the QLOA site. }
  \label{fig_qloa}
\end{figure}

We validate the MRF model using both neutral and convective data of ASL measured at the Qingtu Lake Observation Array (QLOA), as shown in Fig. \ref{fig_qloa}. 
QLOA is a unique field observation station capable of synchronous measurements of the three-dimensional (streamwise, spanwise and wall-normal directions) wind velocity, sand concentration, temperature, humidity, and electric field strength within the three-dimensional ASL turbulent flow. 
High-quality wind data with the highest known friction Reynolds number ($\text{Re}_\tau\sim O(10^6)$) measured at QLOA are validated for ASL studies \cite{wang2016very, Liu_Zheng_2021}. 
Details of QLOA and the ASL data used in this study are provided in Supplemental Material Part B.

For neutral ASL, the streamwise spectrum follows $k_x^{-1}$ and $k_x^{-5/3}$ scalings, with transition scales at $O(z)$ and $O(\delta)$. 
Therefore, we describe the ASL data using the MRF model with $N=2$. 
To avoid spectral errors, we fit the second-order structure function, which is defined in the physical space and exhibits lower error compared to the one-dimensional spectrum. 
Since both the MRF model and ASL data correspond to the same second-order structure function, the resulting model spectrum accurately represents the smoothed spectrum of the ASL data. 
More details about the fitting procedure can be found in Supplemental Material Part A.3. 

\begin{figure}
\includegraphics[width=.47\textwidth]{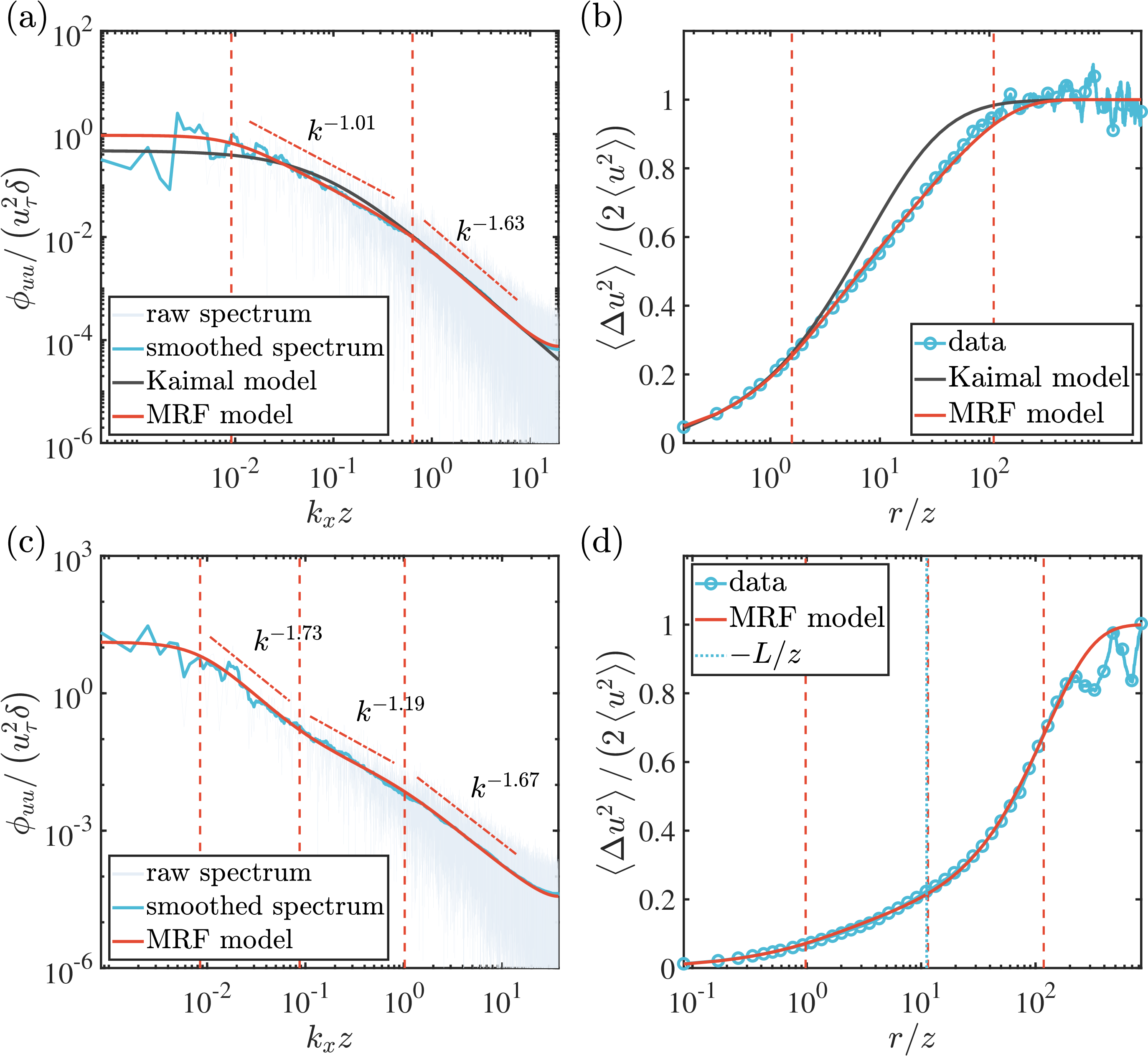}%
\caption{Spectra and second-order structure functions for (a, b) neutral data fitted with the MRF model ($N=2$), and (c, d) convective data fitted with the MRF model ($N=3$). 
(a) Spectrum and (b) second-order structure function of the Kaimal model are presented for comparison.
Vertical orange dashed lines indicate transition wavenumber $\lambda_i z/(U\Delta t)$ and scale $U\Delta t/(\lambda_i z)$. 
The friction Reynolds number is $3.74 \times 10^6$ and the dimensionless height is $u_\tau z/\nu = 4.26 \times 10^4$ for the neutral data.
For the convective case, the friction Reynolds number is $4.19 \times 10^6$, the dimensionless height is $u_\tau z/\nu = 4.77 \times 10^4$, and $z/L = -0.09$.
 \label{fig_MRF_fit}}
\end{figure}

Fig.\ref{fig_MRF_fit} shows examples of using the MRF model to capture the ASL spectrum and second-order structure function in neutral ((a) and (b)) and convective ((c) and (d)) ASL. 
For neutral cases, the MRF model well captures the $k_x^{-1}$ and $k_x^{-5/3}$ scalings at different ranges. 
In contrast, empirical spectral models such as the Kaimal model \cite{kaimal1972spectral} are limited in that they can only represent the small-scale $k_x^{-5/3}$ scaling, failing to accurately describe the behaviour at larger scales. 
For the more complicated convective situation, the MRF model successfully identifies the three spectral ranges, with the transition wavenumbers corresponding approximately to the characteristic scales $z$, $-L$, and $\delta$, which are shown in the Supplemental Material Part C.
The power exponents for the convective-dynamic and dynamic ranges are close to their theoretical values of $-5/3$, while the power exponent for the dynamic range deviates from its theoretical value $-1$. 
This deviation may be due to the insufficient separation between the scales $z$ and $-L$ under convective conditions, as well as the influence of high-order buoyancy terms becoming significant \cite{tong2019multi}.

\section{Analyzing ASL spectrum using MRF model. }--- We applied the MRF model to streamwise velocity, vertical velocity, and temperature, and collected statistical results for key spectral features, including transition scales and power exponents. 
Additionally, the low-wavenumber exponents for vertical velocity and temperature were adjusted based on MOST constraints for variance scaling.
These results are summarized in Table \ref{s0s1s2l0l1l2}, with further details available in Supplemental Material Part C. 

For streamwise velocity at heights with $z < -L$, the streamwise velocity spectrum can be divided into three distinct ranges: the convective-dynamic range ($1/\delta < k_x < -1/L$), characterized by a $k_x^{-5/3}$ scaling; the dynamic range ($-1/L < k_x < 1/L_\epsilon$), with a $k_x^{-1}$ scaling; and the inertial range ($1/L_\epsilon < k_x \ll 1/\eta$) with $k_x^{-5/3}$, where $\eta$ is the Kolmogorov scale. 
Note that for clarity of presentation, we have used $z$ as the characteristic scale in Fig. \ref{fig_MRF_fit}. 
However, due to the control of energy dissipation rate on near-wall turbulence structure \cite{davidson2014universal, Tang2023Similarity}, the appropriate characteristic scale for the streamwise spectrum is $L_\epsilon=u_\tau^3/\epsilon\approx \kappa z(1-z/L)^{-1}$, which is approximately $O(z)$. 
Our subsequent discussion of \eqref{u2_revise_2} further demonstrates the necessity of using $L_\epsilon$ as the characteristic scale for the inertial range. 

As to the vertical velocity and temperature, we retain the three power-law behaviours from the MMO framework with details shown in Table \ref{s0s1s2l0l1l2}.
It suggests that the transition scales and power exponents predicted by MMO need to be modified: The transition scales for both vertical velocity and temperature are observed to be smaller in comparison to those of streamwise velocity. 
For vertical velocity, the transition scale $L_0$ between the inertial and dynamic ranges is $O(0.1z)$, while the transition scale $L_1$ between the dynamic and convective-dynamic ranges is $O(z)$, consistent with findings for canonical boundary-layer turbulence \cite{Mcnaughton2007ScalingPO,yang2017structure}. 
Similar transition scales are also observed for temperature.
Our new finding based on the MRF model is that the largest transition scale for vertical velocity and temperature spectra, $L_2$, is of $O(-L)$, which is distinctively smaller than $O(\delta)$. 
Also, we find that the power exponents $-1$ and $-1/2$ for the vertical velocity in the dynamic range and the convective-dynamic range, respectively, differ from the values of $1$ and $1/3$ proposed by MMO. 
The power exponent of $-4/3$ for temperature in the convective-dynamic range deviates from the $-1/3$ proposed by MMO, which is explained in the below refined MMO section.
These findings update our understanding of the spectral energy distribution of velocity and temperature in convective boundary-layer turbulence.


\begin{table}[htbp]
\centering
\tabcolsep=0.2cm
\renewcommand\arraystretch{1.4}
\begin{tabular}{ccccccc}
\hline\hline
& $S_0$ & $L_0$ & $S_1$ & $L_1$ & $S_2$ & $L_2$ \\
\hline 
$u$ & $-5/3$ & $O(L_\epsilon)$ & $-1$ & $O(-L)$ & $-5/3$ & $O(\delta)$ \\
$w$ & $-5/3$ & $O(0.1z)$ & $-1$ & $O(z)$ & $-1/2$ & $O(-L)$ \\
$\theta$ & $-5/3$ & $O(0.1z)$ & $-1$ & $O(z)$ & $-4/3$ & $O(-L)$ \\
\hline\hline 
\end{tabular}
\caption{Statistical results for power exponents and transition scales. $S_0$, $S_1$, and $S_2$ represent the power exponents for the inertial, dynamic, and convective-dynamic ranges, respectively. $L_0$, $L_1$, and $L_2$ denote the upper bounds of the inertial, dynamic, and convective-dynamic ranges, respectively. }
\label{s0s1s2l0l1l2}
\end{table}

\section{Refined MOST based on MRF model. }---
The simple analytical form of the spectrum of MRF model offers a new way of analyzing and quantifying numerical and observational data by quantifying characteristic scales and exponents.
Here, we apply this idea to refine the MOST theory for streamwise turbulent kinetic 
The dimensional analysis of MOST is based on the assumption that the boundary layer thickness $\delta$ does not directly affect the atmospheric surface layer (ASL). 
The only available dimensionless combination is $z/L$. 
However, very-large-scale motions on the order of $\delta$ also contribute to streamwise velocity fluctuations \cite{wang2016very}. 
Thus, we introduce $\delta$ as a characteristic length scale for streamwise velocity \cite{tong2015multipoint}.

We examine the consistency between velocity variances and spectra, given that variance corresponds to the integral of the one-dimensional spectrum. 
To determine the relationship between variance, transition scales, and spectral exponents, we use a simplified spectral model as described in \citet{vassilicos2015streamwise}. 
The spectrum is divided into four ranges: (i) a plateau for $k_x < 1/L_2$; (ii) a low-wavenumber scaling of $k_x^{-m}$ for $1/L_2 < k_x < 1/L_1$; (iii) a mid-wavenumber scaling of $k^{-1}$ for $1/L_1 < k < 1/L_0$; and (iv) a high-wavenumber scaling of $k^{-5/3}$ for $k > 1/L_0$ in the inertial subrange. 
By matching the leading order of the spectrum, we determine the spectral coefficients, and integrating the spectrum over $k_x$ yields:
\begin{multline}\label{u2_MMO}
    \mean{u^{+2}} = C_0 \left(\frac{L_0}{L_\epsilon}\right)^{2/3}
    \left[\frac{3}{2} + \frac{1}{1-m} \right. \\
    \left. - \frac{m}{1-m}\left(\frac{L_1}{L_2}\right)^{1-m}
    + \ln \left(\frac{L_1}{L_0}\right) \right]. 
\end{multline}
With $L_0 \sim L_\epsilon$, $L_1 \sim -L$, $L_2 \sim \delta$, and $m = -5/3$ \cite{tong2015multipoint}
, we obtain
\begin{equation}\label{u2_revise_2}
    \mean{u^{+2}}=A_u \left(-\frac{\delta}{L}\right)^{2/3} 
    +B_u \ln \left(1-\frac{L}{z}\right)+C_u. 
\end{equation}
where $A_u$, $B_u$, and $C_u$ are constants to be determined, and $L_\epsilon\approx\kappa z(1-z/L)^{-1}$ is used. 
For strong convective conditions, where $L_\epsilon \approx -L$, Eq. \eqref{u2_revise_2} simplifies to a $2/3$ power function of $-\delta/L$. 

\begin{figure*}
\includegraphics[width=\textwidth]{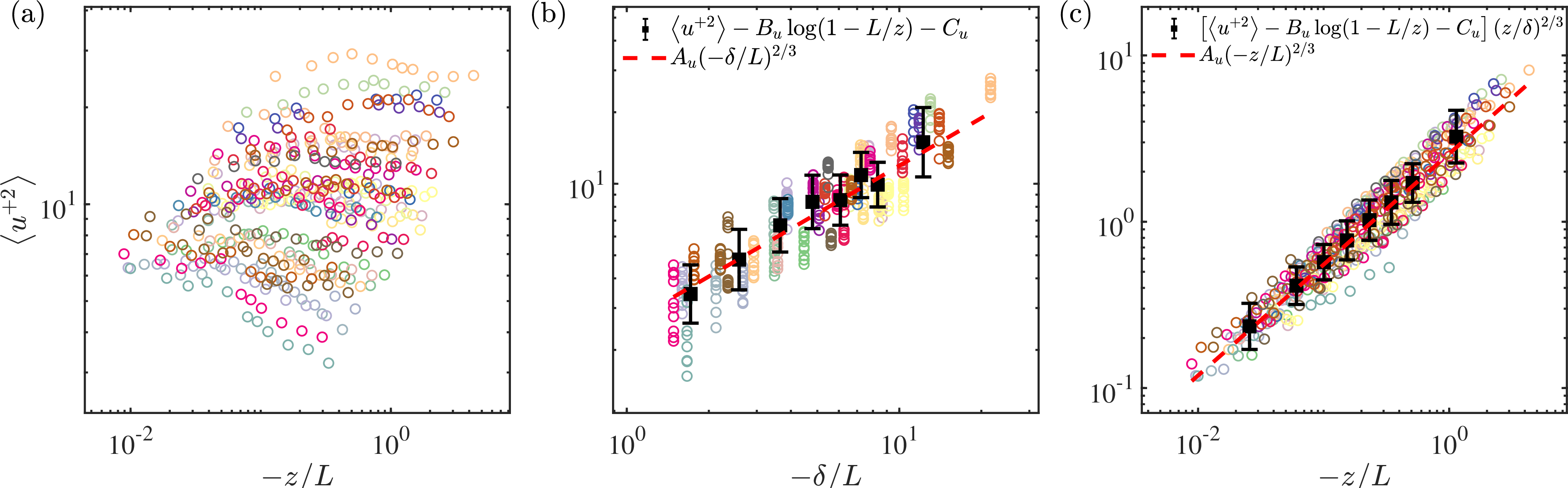}%
\caption{
Comparison of MOST with the expression Eq. \eqref{u2_revise_2} for $\mean{u^{+2}}$ from ASL data. (a) Non-scaling result of MOST for $\mean{u^{+2}}$. (b) and (c) are plots of Eq. \eqref{u2_revise_2}.\label{fig_u2}}
\end{figure*}

Fig. \ref{fig_u2} shows the compensated form of Eq. \eqref{u2_revise_2}, with $A_u = 2.57$, $B_u = 0.33$, and $C_u = 1.19$, which are obtained as the average of the fitting results. 
Compared to the uncollapsed MOST result in MOST, the new formulation involving $-\delta/L$ effectively captures the power-law scaling of $\mean{u^{+2}}$. 
Furthermore, with the empirically determined values for $A_u$, $B_u$, and $C_u$, Eq. \eqref{u2_revise_2} accurately captures the observed variations in $\mean{u^{+2}}$. 
Recently, \citet{stiperski2023generalizing} proposed an extension of MOST by introducing a multiplicative term that quantifies turbulence anisotropy. 
When using $z/L$ as the independent variable, our expression \eqref{u2_revise_2} also accounts for the anisotropic effects, quantized by $(z/\delta)^{2/3}$, which is shown in Fig. \ref{fig_u2}(c). 
However, \eqref{u2_revise_2} offers clearer interpretability.

For vertical velocity and temperature, we use the asymptotic behaviours $\mean{w^{+2}} \sim \left(-z/L\right)^{2/3}$ and $\mean{\theta^{+2}} \sim \left(-z/L\right)^{-2/3}$ from MOST as constraints to determine the low-wavenumber scaling. 
More details are available in Supplemental Material Part D.

\section{Refined MMO based on MRF model.}--- 
As listed in Table \ref{s0s1s2l0l1l2}, differences in transition scales and power exponents for vertical velocity and temperature between the ASL data and MMO are observed. 
Here, we focus on explaining the exponent $-4/3$ in the convective-dynamic range of the temperature spectrum.
MOST predicts that $\mean{\theta^{+2}} \sim (-z/L)^{-2/3}$.
On the other hand, by integrating the temperature spectrum obtained from the MRF model, we obtain
\begin{equation}\label{theta2_MMO}
\begin{aligned}
        \mean{\theta^{+2}}&=
    C''_0 \left(-\frac{z}{L}\right)^{-1/3}
    \left[\frac{3}{2}+\frac{1}{1-m}\right.
    \\&
    \left.
    -\frac{m}{1-m}\left(\frac{z}{L}\right)^{1-m}
    +\ln \left(\frac{L_1}{L_0}\right)
    \right] \sim \left(\frac{z}{L}\right)^{2/3-m},
\end{aligned}
\end{equation}
where the limit of strong convection is considered, and $m$ is the scaling exponents in the convective-dynamic range.
Thus, the consistency between MOST and (\ref{theta2_MMO}) implies $m=-4/3$, which is consistent with the ASL data. 
Further details are provided in Supplemental Material Part D, where the bound for the scaling exponent of the convective-dynamic range of the vertical velocity spectrum is also obtained.



\section{Conclusions. }---
ASL turbulence exhibits multiscale correlations in both the streamwise and vertical directions, making traditional models with a single correlation parameter insufficient. 
To address this, we propose a multiscale dynamical model inspired by tempered fractional Brownian motion--the MRF model--featuring broader applicability. 
The MRF model incorporates characteristic scales, capturing both streamwise and vertical information, as well as power exponents representing correlations or fractal dimensions across scales.
Using extensive field-measurement ASL data from QLOA, we apply the MRF model, obtain statistical results, and validate the proposed theory and model. 
Building on the insights from Monin-Obukhov similarity theory (MOST), which characterizes vertical scales, and multi-point Monin-Obukhov theory (MMO), which emphasizes two-point statistics in the streamwise wavenumber space, we apply the MRF model to convective ASL, effectively bridging MOST and MMO through the relationship between the one-dimensional spectrum and variance, which is possible only when the transition scales and power exponents are quantified. 
Leveraging the scaling relationships proposed by MMO, we derive a new expression for $\mean{u^{+2}}$ (cf. Eq. \eqref{u2_revise_2}). 
Additionally, using the asymptotic scaling constraints from MOST for $\mean{w^{+2}}$ and $\mean{\theta^{+2}}$, we refine the low-wavenumber scalings in MMO. 
The complete spectral model for streamwise velocity, vertical velocity, and temperature is summarized in Table \ref{s0s1s2l0l1l2}.
This method provides a stochastic representation based on the composition of fractional Brownian motion with different Hurst exponents obtained from second-order structure function, which can be empirically obtained with much less error compared with the spectrum. 
In contrast to prior approaches, the MRF model captures multiscale and multi-effect behaviours, which cannot be described by the single Hurst index used in traditional statistical models, providing a novel framework for understanding complex systems. 
Furthermore, using the MRF model, we can generate synthetic turbulent data of ASL by capturing the multi-range and multiscale behaviour. 

\begin{acknowledgments}
F-CZ and J-HX acknowledge financial support from the National Natural Science Foundation of China, grant numbers 12272006, 12472219 and 42361144844, and from the Laoshan Laboratory under grant numbers LSKJ202202000, LSKJ202300100.
XZ acknowledges financial support from the National Natural Science Foundation of China, grant numbers 92052202 and 12388101.
\end{acknowledgments}

\bibliography{bib.bib}
\end{document}


\title{Supplemental Material: Multi-range fractional model for convective atmospheric surface-layer turbulence}

\author{Fei-Chi Zhang}
\affiliation{Department of Mechanics and Engineering Science at College of Engineering, and State Key Laboratory for Turbulence and Complex Systems, Peking University, Beijing 100871, P. R. China}
\author{Jin-Han Xie}
\email{jinhanxie@pku.edu.cn}
\affiliation{Department of Mechanics and Engineering Science at College of Engineering, and State Key Laboratory for Turbulence and Complex Systems, Peking University, Beijing 100871, P. R. China}
\author{Xiaojing Zheng}
\email{xjzheng@xidian.edu.cn}
\affiliation{Research Center for Applied Mechanics, Xidian University, Xi’an 710071, P. R. China}

\date{\today}
\maketitle

\section{MRF model}
\subsection{Introducting MRF model}
The MRF is a extension of tempered fractionally integrated (TFI) model, whose expression is \cite{giraitis2000stationary,meerschaert2014tempered} 
\begin{equation}\label{ARTFIMA}
	(1-\ee^{-\lambda} \mathcal{B})^d u_t =\epsilon_t, 
\end{equation}
where $\lambda>0$. 
Tempered fractional calculus is helpful for characterizing non-local turbulence structures in large-eddy simulation \cite{samiee2022tempered}, and its basic properties can be found in \citet{sabzikar2015tempered,sabzikar2018invariance}. 
In contrast to the fractional model, the autocorrelation of tempered fractionally integrated (TFI) model Eq.\eqref{ARTFIMA} is integrable, and its spectrum follows power law at large wavenumber, which makes Eq.\eqref{ARTFIMA} mathematically more tractable and applicable in many cases, especially for $d=5/6$ applied to Kolmogorov turbulence \cite[cf.][]{mandelbrot1968fractional}.
The TFI model represents a discrete analogue of tempered fractional Brownian motion \cite{sabzikar2018invariance} with a spectrum
\begin{equation}\label{ARTFIMA_psd}
    f(k)= \frac{\sigma_\epsilon^2}{2\pi} 
    \left(1-2\ee^{-\lambda}\cos k+\ee^{-2\lambda}\right)^{-d},\,-\pi\leq k\leq \pi, 
\end{equation}
where $\sigma_\epsilon^2$ is the variance of $\epsilon_t$ and $k$ is the nondimensional frequency. 
As $k/\lambda\to 0$, $f$ tends to a constant $\sigma_\epsilon^2 \left(1-\ee^{-\lambda}\right)^{-2d}/(2\pi)$, and when $k\gg \lambda$, $f$ behaves as
\begin{align}\label{ARTFIMA_psd_power}
    f(k)\approx& \frac{\sigma_\epsilon^2}{2\pi} 
    \left[1-
    2\left(1-\lambda+\frac{\lambda^2}{2}\right)
    \left(1-\frac{k^2}{2}\right)
    +1-2\lambda+2\lambda^2\right]^{-d}\nonumber\\
    \approx& \frac{\sigma_\epsilon^2}{2\pi} 
    \left(k^2+\lambda^2\right)^{-d}\sim k^{-2d}. 
\end{align}
Thus TFI model is applicable to the cases where the spectrum is flat at small wavenumber and has power-function decay at large wavenumber. 
However, Eq.\eqref{ARTFIMA} which captures one scaling, is not applicable to the spectrum of ASL turbulence with multiple scalings.
So, we extend the TFI model to the MRF model by composing TFI models: 
\begin{equation}\label{MRF}
	\prod \limits_{i=1}^{N}
	 (1-\ee^{-\lambda_i} \mathcal{B})^{d_i-d_{i-1}} 
	 u_t =\epsilon_t, 
\end{equation}
where $d_0=0$, $N$ is the number of characteristic length scales of the system, and $\lambda_i>0$.
The model's spectrum is
\begin{align}\label{MRF_spectrum}
	f(k)= \frac{\sigma_\epsilon^2}{2\pi} 
	\prod_{i=1}^{N}\left(1-2\ee^{-\lambda_i}\cos k+\ee^{-2\lambda_i}\right)^{-(d_i-d_{i-1})},\,-\pi\leq k\leq \pi.  
\end{align}
With multiple tempered fractional operators, the MRF model provides a global expression. 
The product of multiple tempered fractional operators leads to multiple power-law behaviors of the model spectrum. 
One advantage of the MRF model is its linear form that permits analytical expressions. 
Another advantage is that the model parameters have clear physical meanings: the parameter $\lambda_i$ is the transition wavenumber of different spectral regimes, and the parameter $d_i$ corresponds to the power exponent of the spectrum approximately. 
\subsection{Properties of MRF model}
The expression for the linear model is
\begin{equation}
    \mathcal{L} (\mathcal{B}) u_t =\epsilon_t, 
\end{equation}
where $\mathcal{L}$ is a function of the lag operator $\mathcal{B}$ ($\mathcal{B}u_t=u_{t-1}$) and $\epsilon_t$ is an uncorrelated random variable that usually follows Gaussian distribution for simplicity. 
If the model is invertible the expression can be converted into
\begin{equation}
     u_t =\Psi(\mathcal{B})\epsilon_t.  
\end{equation}
Since the roots of $\mathcal{L}(z)$ of the TFI model avoid the unit circle \cite{sabzikar2019parameter} and the $\mathcal{L}(z)$ of MRF model is the cumulative product of $\mathcal{L}(z)$ of TFI model, the MRF model is stationarity. 

For the TFI model, the tempered fractional integration operator in Eq.\eqref{ARTFIMA} can be expanded as \cite{meerschaert2014tempered, sabzikar2018invariance}
\begin{equation}
    (1-\ee^{-\lambda} \mathcal{B})^{-d}=\sum_{j=0}^{\infty} \frac{\Gamma(j+d)\ee^{-\lambda j}\mathcal{B}^j}{\Gamma(j+1)\Gamma(d)}, 
\end{equation}
where $\Gamma(\cdot)$ is the Gamma function. 
Then we have
\begin{equation}\label{MRF_expansion1}
	\prod \limits_{i=1}^{N}
	 (1-\ee^{-\lambda_i} \mathcal{B})^{-(d_i-d_{i-1})}= \sum_{j=0}^{\infty} \psi_j\mathcal{B}^j, 
\end{equation}
and 
\begin{equation}
	\psi_j=\sum_{\substack{j_1+j_2+\cdots j_N=j\\j_1,j_2,\cdots,j_N\geq0}}
    \frac{\Gamma(j_1+d_1)\ee^{-\lambda_1 j_1}}{\Gamma(d_1)\Gamma(j_1+1)}
    \frac{\Gamma(j_2+d_2-d_1)\ee^{-\lambda_2 j_2}}{\Gamma(d_2-d_1)\Gamma(j_2+1)}\cdots \frac{\Gamma(j_N+d_N-d_{N-1})\ee^{-\lambda_N j_N}}{\Gamma(d_N-d_{N-1})\Gamma(j_N+1)}, 
\end{equation}
where $i$, $j$, $j_1$ to $j_N$ used are all integers. 
Since the TFI model satisfies
\begin{equation}
    \sum_{j=0}^{\infty} \left|
    \frac{\Gamma(j+d)\ee^{-\lambda j}}{\Gamma(d)\Gamma(j+1)}
    \right|<\infty
\end{equation}
for all $\lambda>0$ and $d \notin \mathbb{Z}$, implying the causality. 
Then we obtain
\begin{align}
    \sum_{j=0}^{\infty} \left|\psi_j\right|&= 
    \sum_{j=0}^{\infty} \left| 
    \sum_{\substack{j_1+j_2+\cdots j_N=j\\j_1,j_2,\cdots,j_N\geq0}}
    \frac{\Gamma(j_1+d_1)\ee^{-\lambda_1 j_1}}{\Gamma(d_1)\Gamma(j_1+1)}
    \frac{\Gamma(j_2+d_2-d_1)\ee^{-\lambda_2 j_2}}{\Gamma(d_2-d_1)\Gamma(j_2+1)}\cdots \frac{\Gamma(j_N+d_N-d_{N-1})\ee^{-\lambda_N j_N}}{\Gamma(d_N-d_{N-1})\Gamma(j_N+1)}
    \right|    \nonumber\\ 
    &\leq \sum_{j=0}^{\infty} \sum_{\substack{j_1+j_2+\cdots j_N=j\\j_1,j_2,\cdots,j_N\geq0}}
    \left| \frac{\Gamma(j_1+d_1)\ee^{-\lambda_1 j_1}}{\Gamma(d_1)\Gamma(j_1+1)} \right|
    \left| \frac{\Gamma(j_2+d_2-d_1)\ee^{-\lambda_2 j_2}}{\Gamma(d_2-d_1)\Gamma(j_2+1)} \right|
    \cdots
    \left| \frac{\Gamma(j_N+d_N-d_{N-1})\ee^{-\lambda_N j_N}}{\Gamma(d_N-d_{N-1})\Gamma(j_N+1)} \right| \nonumber\\ 
    &\leq \prod_{i=1}^{N}
    \left(
    \sum_{j=0}^{\infty} 
    \left| \frac{\Gamma(j+d_i-d_{i-1})\ee^{-\lambda_i j}}{\Gamma(d_i-d_{i-1})\Gamma(j+1)} \right|
    \right)
    <\infty. 
\end{align}
Thus MRF model is casual. 
Following a similar process, it can be proved that the MRF model is invertible since the TFI model is invertible \cite{sabzikar2018invariance}.

Eq.\eqref{MRF} can be expressed as
\begin{equation}\label{MRF_expansion2}
    u_t=\sum_{j=0}^{\infty}\psi_j \epsilon_{t-j}, 
\end{equation}
and thus the model's autocorrelation is
\begin{equation}
    \mean{u_t u_{t-i}}=\sum_{j=0}^{\infty}\psi_j \psi_{j+i} \sigma_\epsilon^2. 
\end{equation}
According to the definition of the spectral density of the linear process \cite{shumway2000time}, the spectrum of the MRF model is
\begin{equation}
    f(k)=\left| \Psi(\ee^{-\ii k})\right|^2 f_\epsilon (k), 
\end{equation}
where $\ii$ is imaginary unit and $f_\epsilon (k)=\sigma_\epsilon^2/(2\pi)$. 
Then the spectrum of MRF model is obtained, see Eq.\eqref{MRF_spectrum}. 
And the autocorrelation of the MRF model can also be obtained from its spectrum
\begin{equation}\label{MRF_fourier}
    \mean{u_t u_{t-i}} = \int_{0}^{\pi} f(k) \cos (ki)\, \dif k. 
\end{equation}
\subsection{Fitting procedure of MRF model}
For a velocity series $\{u_t\}$, we can calculate the spectrum $\phi_{uu}(k_x)$, autocorrelation $R_{uu}(r)=\mean{u_{x} u_{x+r}}/\mean{u^2}$, and the second-order structure function
\begin{equation}
    \mean{\Delta u^2(r)} =2 \mean{u^2} \left( 1- R_{uu}(r)\right), 
\end{equation}
where $k_x$ is the wavenumber, $x$ is the spatial location to be averaged, $r$ is the spatial scale that measures the distance between two points, and $\mean{u^2}$ is the variance of $\{u_t\}$. 
Then, we can fit the MRF model with data's spectrum or autocorrelation. 
Traditional methods for parameter estimation of the TFI model include Whittle likelihood-based estimation technique \cite{sabzikar2019parameter} and the spectrum-based nonlinear least square technique \cite{bhootna2023gartfima}. 
However, considering that the spectrum of ASL data contains relatively large errors, in this work we use another fitting method based on the second-order structure function. 
Compared to the spectrum, the structure function has small errors at inertial and dynamic ranges, thus making it more convenient to estimate the errors. 
According to Eq.\eqref{MRF_fourier} we get
\begin{equation}
    \tilde R_{uu}(r)=\frac{\int_{0}^{\pi} f(k) \cos \left(k\frac{r}{U\Delta t}\right)\, \dif k}{\int_{0}^{\pi} f(k)\, \dif k},
\end{equation}
where $\tilde R_{uu}$ is the autocorrelation of MRF model. 
The fitting error is defined as
\begin{equation}\label{MRF_error}
    \mathcal{E}(\bm{d},\bm{\lambda})=\int_{r_{\min}}^{r_{\max}}
    \frac{\left|\tilde R_{uu}-R_{uu}\right|}{1-R_{uu}}
    \,\dif \ln\left(\frac{r}{z}\right), 
\end{equation}
where $\bm{d}=(d_1, d_2, \cdots, d_N)$, $\bm{\lambda}=(\lambda_1, \lambda_2, \cdots, \lambda_N)$, and $r_{\min}$ and $r_{\max}$ are the minimum and maximum scales for evaluating the error of the structure function, respectively. 
Since $R_{uu}$ oscillates around zero for large scales and since $R_{uu}<1$ for $r > 0$, the denominator in Eq.\eqref{MRF_error} takes $1 - R_{uu}$ instead of $R_{uu}$. 
A nonlinear optimization method is then adopted to solve for the parameters $\bm{d}$ and $\bm{\lambda}$ corresponding to the smallest $\mathcal{E}$, e.g., the ``fmincon'' or ``fminsearch'' functions in MATLAB and ``GenSA'' package in R. 

Some relationships between model quantities and turbulence quantities are presented as follows. 
With the help of Taylor's frozen hypothesis, the conversion between nondimensional frequency $k$ of the MRF model and  streamwise wavenumber $k_x$ is
\begin{equation}\label{conver_k}
    k_x=\frac{k}{U\Delta t}, 
\end{equation}
where $U$ is the mean velocity and $\Delta t$ is the sampling interval of the series $\{u_t\}$. 
And the conversion between model's spectrum $f$ and turbulent spectrum $\phi_{uu}$ is
\begin{equation}\label{conver_phi}
    \phi_{uu}=f U\Delta t. 
\end{equation}
With these relations, one can obtain the turbulent spectrum from parameters of the MRF model. 

With the fitted model parameters, the transition wavenumbers are obtained as $\lambda_i/\left(U\Delta t\right)$, with $i=1$ to $N$ representing $\lambda_i$ arranged from large to small. 
And the power exponents can be obtained as $2d_i$ (cf. Eq.\eqref{ARTFIMA_psd_power}). 
The second-order expansion at $k = 0$ used in Eq.\eqref{ARTFIMA_psd_power} may be biased for $k > 0$. 
A more accurate method for obtaining power exponents is to perform a power fit of the model spectrum between two adjacent transition wavenumbers, which does not introduce large errors in the value of power exponents since the model spectrum is smooth. 

\section{ASL data and pretreatment}
The ASL data used in this work comes from the Qingtu lake observation array (QLOA). 
QLOA is built on the flat dry lakebed of Qingtu Lake in western China (E: $103^\circ 40^\prime 03^{\prime \prime}$, N: $39^\circ 12^\prime 27^{\prime \prime}$) and the measured ASL data reaches the highest order of magnitude friction Reynolds number ($\sim O(10^6)$) to date. 
Multi-filed quantities are measured simultaneously, including three-dimensional turbulent velocities, temperature, humidity, PM10 concentration, and electric field. 
And the sonic anemometers are connected to data acquisition instruments synchronized with GPS to ensure data synchronization. 
The data used in this study were obtained from the main tower, whose height lies roughly in the logarithmic region of the atmospheric boundary layer. 
The three-component sonic anemometers (Campbell scientific, CSAT-3B) perform the measurements of velocities and temperature synchronously, with a sampling frequency of 50 Hz. 
These anemometers on the main tower are mounted at 0.9, 1.71, 2.5, 3.49, 5, 7.15, 8.5, 10.24, 14.65, 20.96, and 30 m. 
The high-quality data of clear-air and sand-laden ASL flows proved suitable for turbulent boundary layer studies \citep{wang2016very,liu2019amplitude,wang2020large, liu2021investigation}. 

The nonstationary nature of atmospheric turbulence makes it necessary to select and preprocess the raw data. 
The pretreatment procedures are consistent with previous ASL turbulence studies using QLOA data \cite{wang2016very,liu2021investigation, liu2023amplitude}. 
The pretreatments of the one-hour raw data include wind direction adjustment, de-trending, and
stationary wind selection. 
Wind direction changes during field measurements, so it is necessary to adjust raw data to obtain streamwise and spanwise velocities: 
\begin{equation}
	u = u_{ m } \cos\omega + v_{ m } \sin\omega, \quad v = v_{ m } \cos\omega - u_{ m } \sin\omega, 
\end{equation}
where $u_{ m }$ and $v_{ m }$ are measured streamwise and spanwise velocities, $\omega$ is the angle between the actual wind direction and the streamwise direction of QLOA, and $u$ and $v$ are the adjusted streamwise and spanwise velocities, respectively. 
After adjusting the wind direction, de-trending with a low-pass filter with a cutoff wavelength of 20$\delta$ is also required to remove large-scale synoptic signals. 
Since we focus on stationary turbulence, a nonstationary index $\gamma$ is used to judge the stationarity, which is defined as
\begin{equation}
	\gamma = | ( \sigma _ { \text{M} } - \sigma_ { \text{I} } ) / \sigma _ { \text{I} } | \times 100 \%, 
\end{equation}
where $\sigma_{ \text{M} }=\sum_{i=1}^{12}\sigma_i/12$, $\sigma_1, \sigma_2, \cdots, \sigma_{12}$ are the streamwise velocity variances of one-twelfth part of the entire time interval, and $\sigma_{\text{I}}$ is the variance of the overall time interval. 
If the ASL data satisfies $\gamma<30\%$ then it can be regarded as stationary, and we choose the half-hour interval with the smallest $\gamma$ in one hour. 
Other quantities used here are defined as follows. 
The Obukhov length $L$ is calculated at $z=1.71$ m. 
The friction velocity $u_\tau$ is evaluated by averaging $\left(-\mean{uw}\right)^{1/2}$ at three heights, 0.9, 1.71 and 2.5 m. 
Air kinematic viscosity $\nu$ is calculated from the average temperature at standard atmospheric pressure. 
To assess the frictional Reynolds number Re$_\tau$, the boundary layer thickness $\delta$ is estimated to be 150 m, approximately the mean value measured by radar at QLOA \cite{liu2023amplitude}. 
Additionally, we use the Minnesota relationship to estimate $\delta$, which defines it as a constant multiple of the wavelength corresponding to the maximum of the pre-multiplied streamwise spectrum at the highest measurement point, as used in \citet{Mcnaughton2007ScalingPO}. 
We find that using this alternative estimation method for $\delta$ does not affect the current results.
Taylor's frozen hypothesis is used to convert temporal data into spatial data.

\section{Statistical results of MRF model}
\begin{figure}[htbp]
    \centering
    \includegraphics[width=1\textwidth]{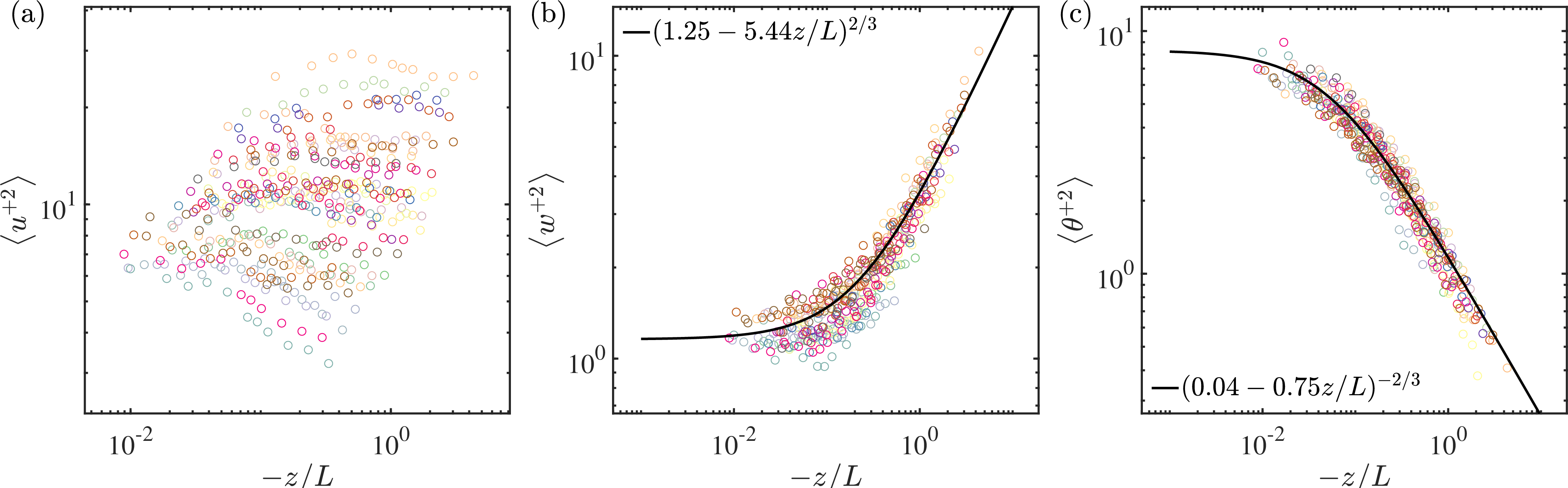}
    \caption{The variances of (a) horizontal velocity, (b) vertical velocity, and (c) temperature from QLOA data. Symbols of the same color indicate different heights for the same set of data. The legends show empirical fitting result of MOST. }
    \label{fig_data_MOST}
\end{figure}

The variances of ASL data used in this work are shown in figure \ref{fig_data_MOST}. 
The friction Reynolds number is on the order of $10^6$. 
It is clear that for convective ASL, $\mean{w^{+2}}$ and $\mean{\theta^{+2}}$ are in good agreement with MOST, while $\mean{u^{+2}}$ is not. 
The $2/3$ scaling of $\mean{w^{+2}}$ and the $2/3$ scaling of $\mean{\theta^{+2}}$ hold for $-z/L$ approximately greater than $0.3$ and $0.1$, respectively. 

\subsection{Results for streamwise velocity}
\begin{figure}[htbp]
    \centering
    \includegraphics[width=.82\textwidth]{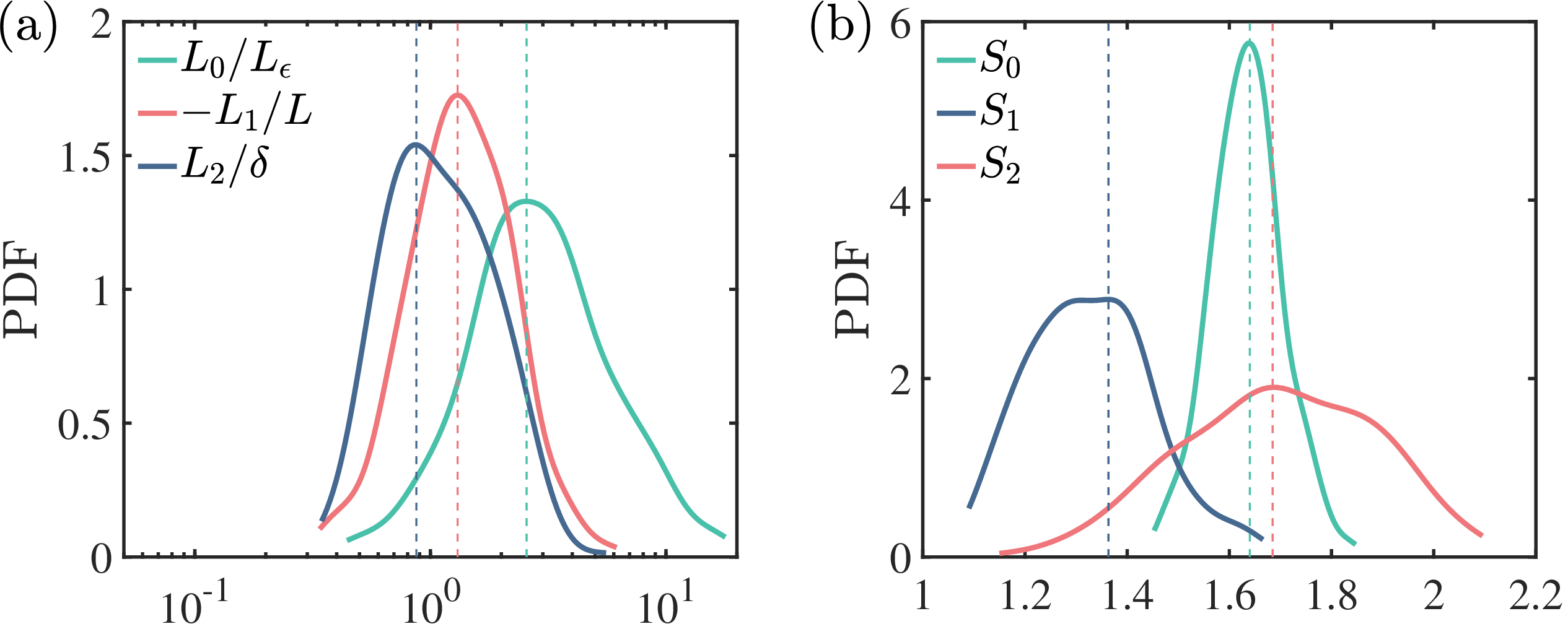}
    \caption{The PDFs of transition scales and power exponents of $\phi_{uu}$. }
    \label{fig_MMO_u_pdf}
\end{figure}

\begin{figure}[htbp]
    \centering
    \includegraphics[width=1\textwidth]{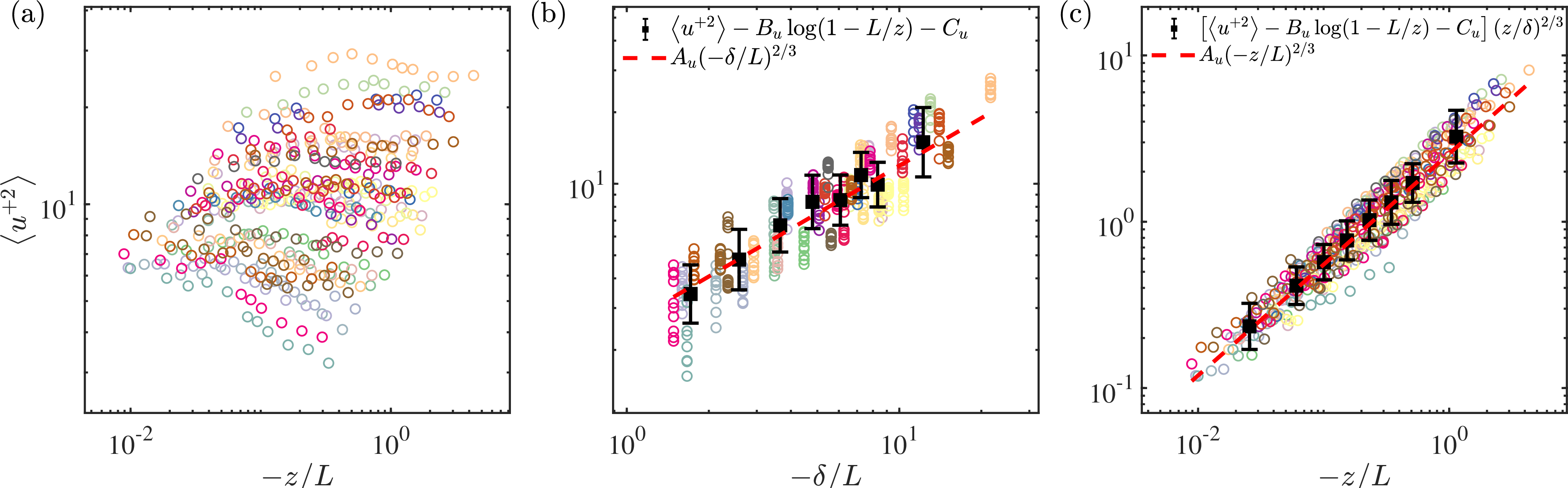}
    \caption{ Validation of the expression \eqref{u2_revise_2} for $\mean{u^{+2}}$ in the ASL data.}
    \label{fig_u2}
\end{figure}

We use the MRF model to extract the transitional wavenumbers and power exponents of $\phi_{uu}$. 
The MRF model obtains spectral information based on the second-order structure function, and thus can discriminate low-wave number spectral information, which is difficult to acquire directly. 
The three length scales $L_0, L_1$, and $L_2$ are depcited in figure \ref{ill_fig_1}. 
For wavenumber ranges of $k>1/L_0$, $1/L_1<k<1/L_0$, and $1/L_2<k<1/L_1$, we denote the power exponents of $\phi_{uu}$ as $S_0$, $S_1$, and $S_2$, respectively.
For convective stratification $-z/L>0.08$, the probability density function (PDF) of transition scales and power exponents are shown in figure \ref{fig_MMO_u_pdf}.
\begin{figure}[htbp]
    \centering
    \includegraphics[width=1\textwidth]{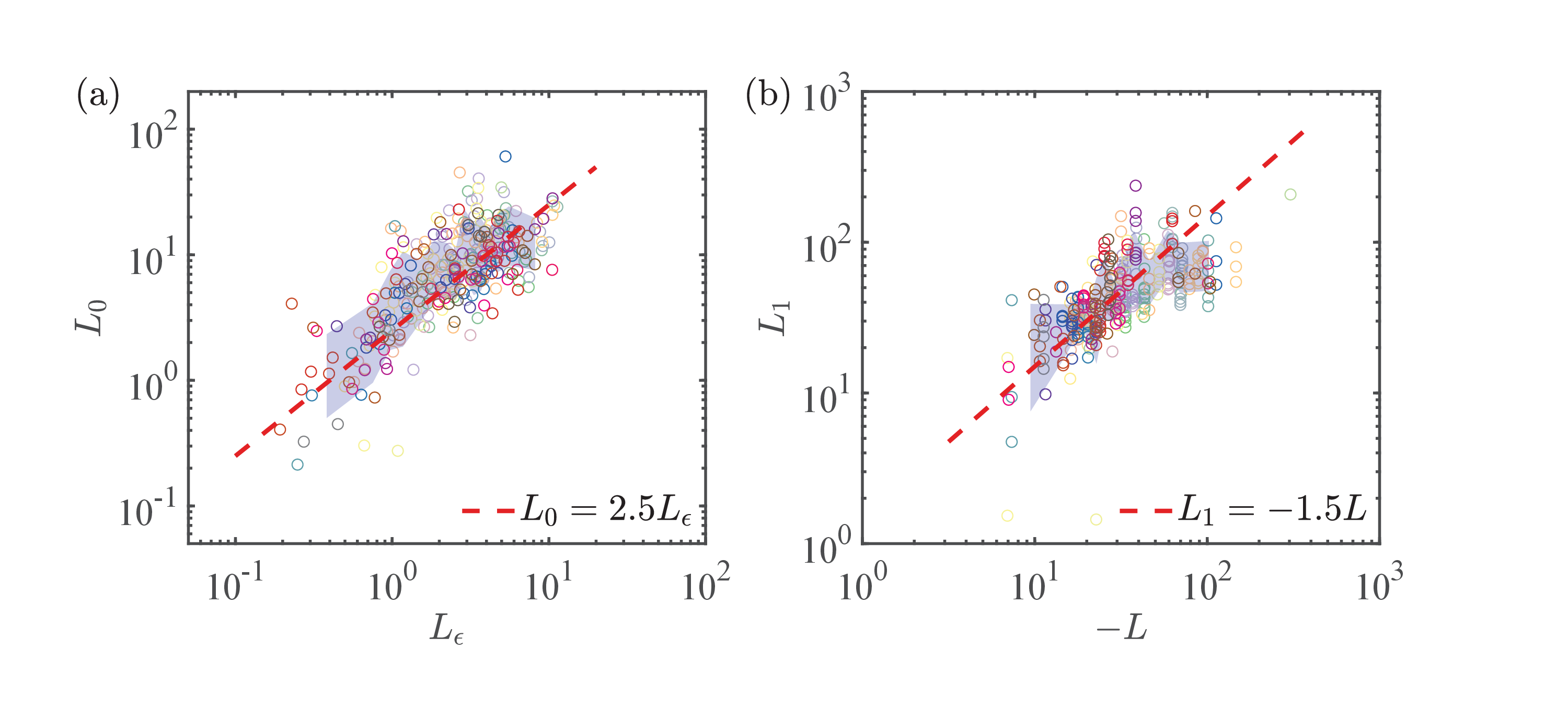}
    \caption{Linear dependence between (a) $L_\epsilon$ and $L_0$, and (b) $-L$ and $L_1$. }
    \label{fig_MMO_u_scale}
\end{figure}
The PDF results show that the transition scales and power exponents fitted by the MRF model are consistent with the predictions of MMO. 
And we check the expression of scales $L_\epsilon/z$ (cf.  \eqref{L_epsilon_relation}) and $L_0/L_1$. 
\begin{figure}[htbp]
    \centering
    \includegraphics[width=1\textwidth]{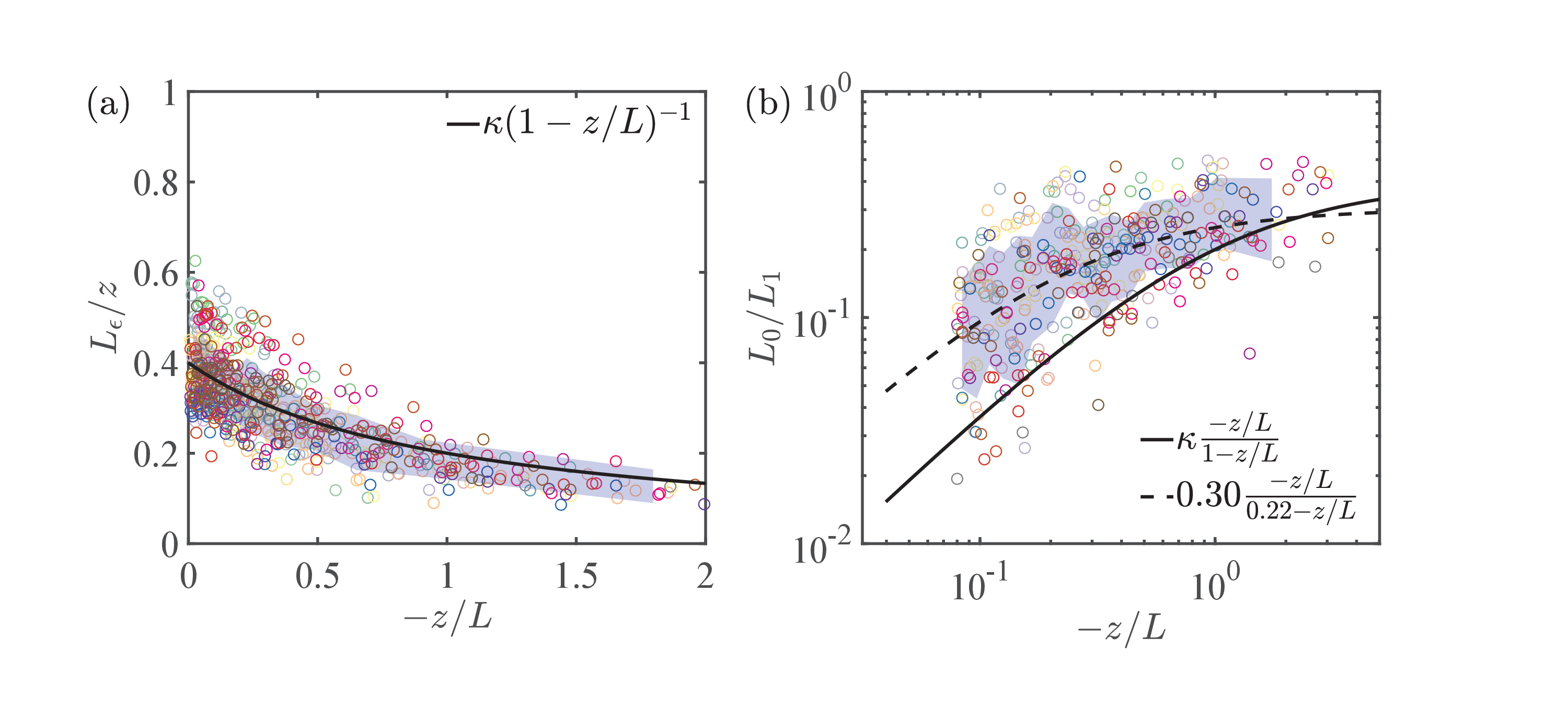}
    \caption{Comparison of expressions of $L_\epsilon/z$ and $L_0/L_1$ with ASL data. }
    \label{fig_MMO_u_L}
\end{figure}
As shown in figure \ref{fig_MMO_u_L} (a), \eqref{L_epsilon_relation} is in good agreement with ASL data, and thus provides an estimation of $L_\epsilon$ in ASL. 
Therefore, with $L_0=L_\epsilon$ and $L_1=-L$, the relation of $L_0$ and $L_1$ should be
\begin{equation}\label{L0_L1}
    \frac{L_0}{L_1}=\kappa \frac{-z/L}{1-z/L}. 
\end{equation}
However, as shown in figure \ref{fig_MMO_u_L} (b), $L_0/L_1$ deviates from \eqref{L0_L1}. 
The curve fitting gives the expression of $L_0/L_1$, as shown in the legend of figure \ref{fig_MMO_u_L} (b). 
However, the behavior common to \eqref{L0_L1} and data is that $L_0/L_1$ scales as $-z/L$ for small $-z/L$, while $L_0/L_1$ tends to a constant for large $-z/L$, since $L_0$ tends to $-L$ as discussed in \S\ref{sec_u}.
These results show that $L_1$ is not exactly equal to $-L$, but is only a rough estimate. 
Data fitting gives the $L_1$ empirical relationship as
\begin{equation}
    L_1=-L\frac{\kappa(0.22-z/L)}{0.3(1-z/L)}. 
\end{equation}
To obtain a concise expression of $\mean{u^{+2}}$, we treat $L_1$ as $-L$, and express \eqref{u2_revise} as
\begin{equation}\label{u2_revise_2}
    \mean{u^{+2}}=A_u \left(-\frac{\delta}{L}\right)^{2/3} 
    +B_u \ln \left(1-\frac{L}{z}\right)+C_u. 
\end{equation}
The constants $A_u$, $B_u$, and $C_u$ are determined from ASL data. 
Figure \ref{fig_u2} shows the compensatory expression of \eqref{u2_revise_2}, where $A_u=3.81$, $B_u=-0.32$, and $C_u=1.57$, with values obtained from the average of the fitting results. 
Compared to figure \ref{fig_data_MOST}(a), the scalings shown in figure \ref{fig_u2}(a) and (b) display clearer power laws. 

\subsection{Results for vertical velocity}
\begin{figure}[htbp]
    \centering
    \includegraphics[width=.82\textwidth]{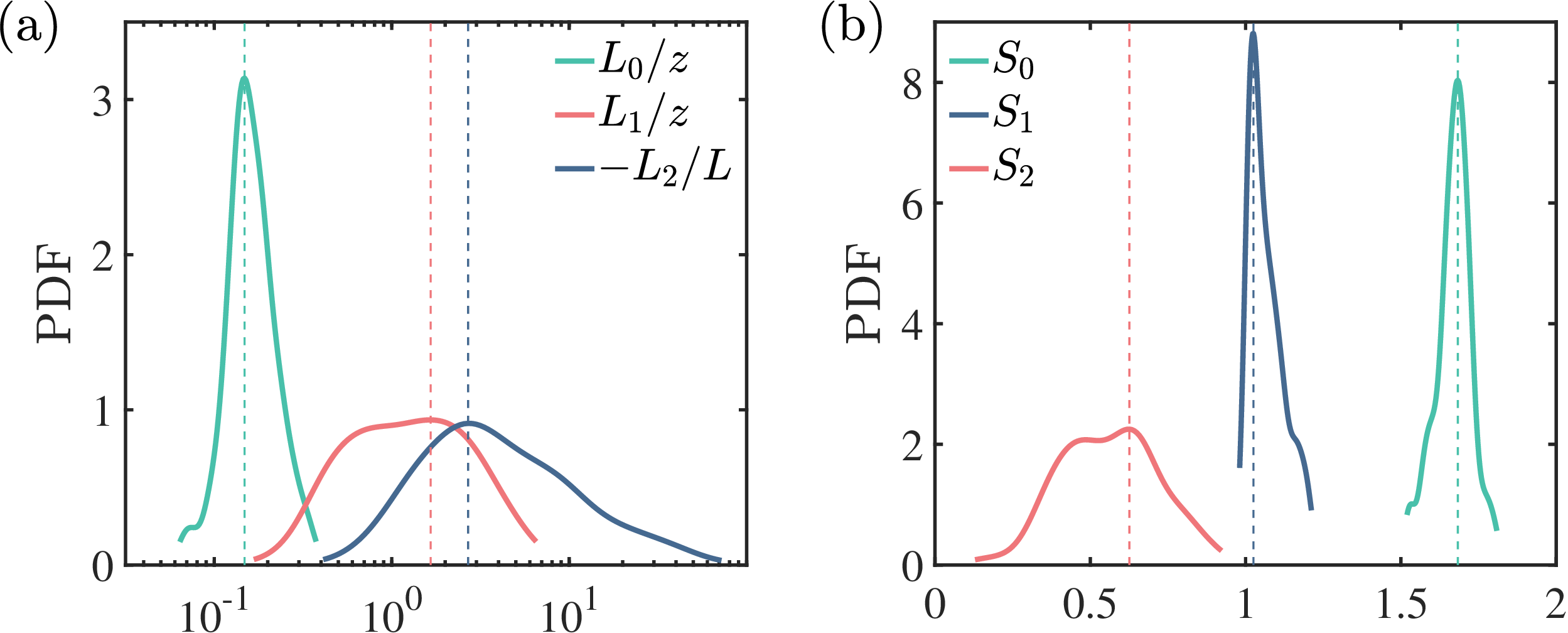}
    \caption{The PDFs of transition scales and power exponents of $\phi_{ww}$. }
    \label{fig_MMO_w_pdf}
\end{figure}

The vertical velocity variance $\mean{w^{+2}}$ has been predicted by MOST. 
So in this section we check the vertical velocity spectrum to explain the behavior of $\mean{w^{+2}}$. 
The three power scalings of $\mean{w^{+2}}$ are investigated through MRF model.
As shown in figure \ref{fig_MMO_w_pdf}(a), the transition scales correspond to $O(0.1z)$, $O(z)$, and $O(-L)$, respectively. 
The power exponents of mid and small scale are close to the theoretical values $1$ and $5/3$. 
The power exponent of large scale is close to $1/2$, which requires further theory. 
\begin{figure}[htbp]
    \centering
    \includegraphics[width=1\textwidth]{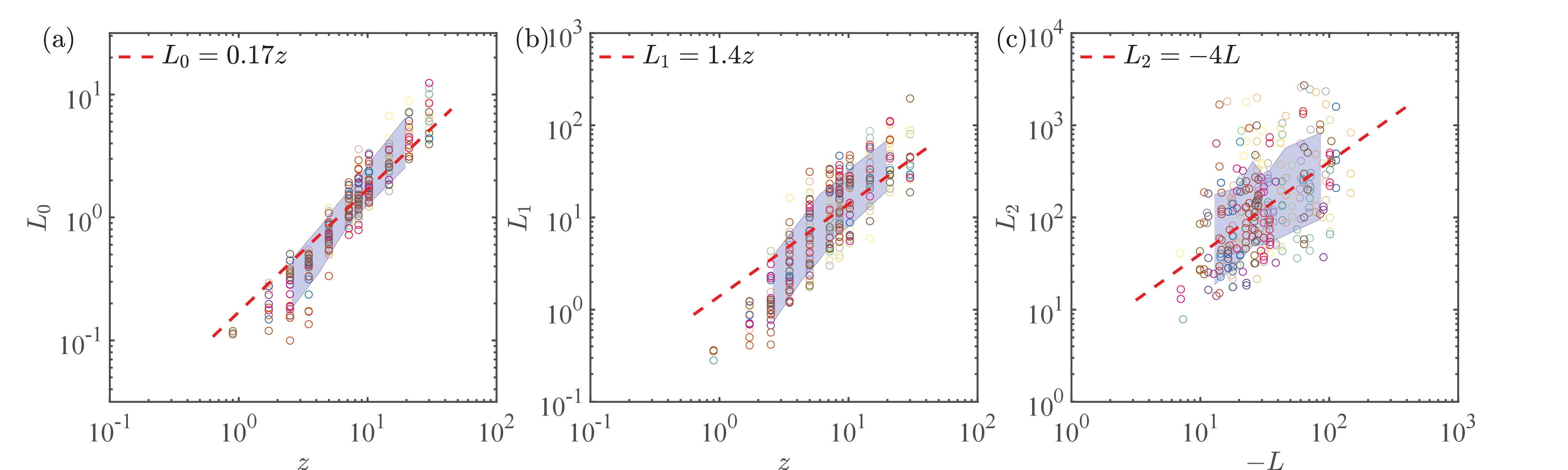}
    \caption{Linear dependence between (a) $z$ and $L_0$, (b) $z$ and $L_1$, and (c) $-L$ and $L_2$. }
    \label{fig_MMO_w_scale}
\end{figure}
With $m=1/2, L_1/L_2=0.1$, the value of term $m(L_1/L_2)^{1-m}/(1-m)$ is about 0.3, which can be neglected compared to $3/2+1/(1-m)$.  
The term $L_1/L_0$ has a weak dependence on $z/L$. 
Therefore, the behavior of $\mean{w^{+2}}$ is mainly determined by the small scale $L_0/L_\epsilon$, which is the coefficient of spectrum (cf. \eqref{coef_spectrum}). 

\subsection{Results for temperature}
\begin{figure}[htbp]
    \centering
    \includegraphics[width=.82\textwidth]{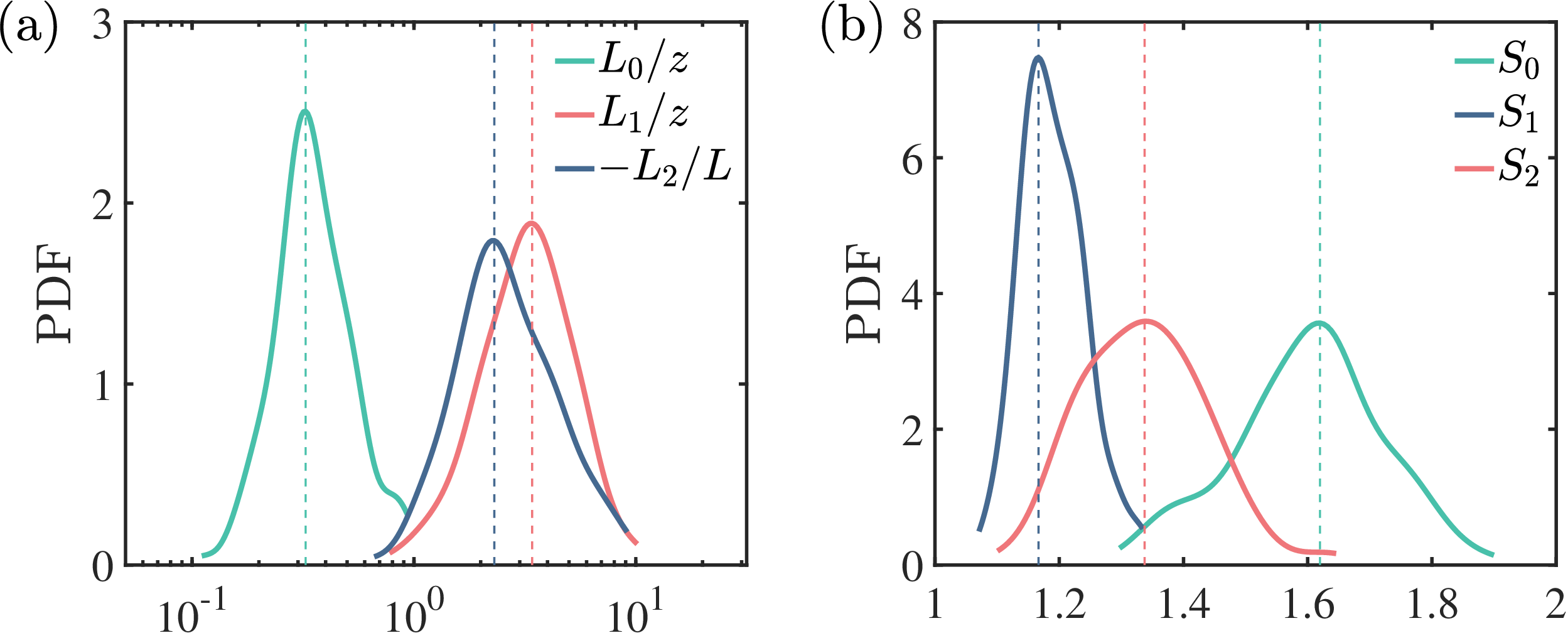}
    \caption{The PDFs of transition scales and power exponents of $\phi_{\theta\theta}$. }
    \label{fig_MMO_theta_pdf}
\end{figure}
\begin{figure}[htbp]
    \centering
    \includegraphics[width=1\textwidth]{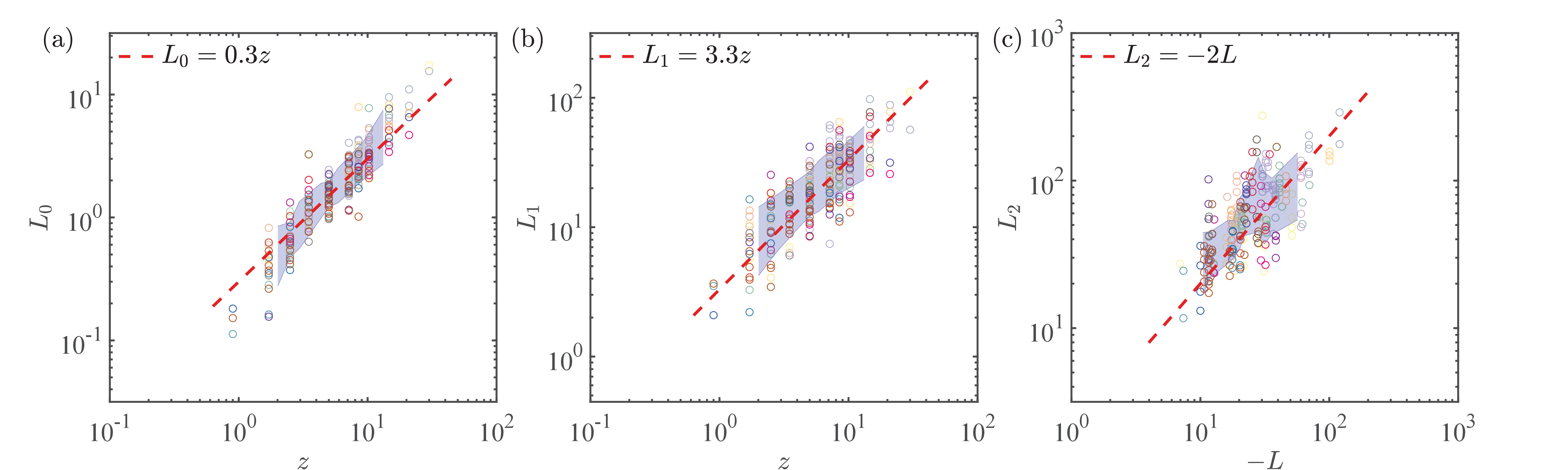}
    \caption{Linear dependence between (a) $z$ and $L_0$, (b) $z$ and $L_1$, and (c) $-L$ and $L_2$. }
    \label{fig_MMO_theta_scale}
\end{figure}
MOST predicts the $2/3$ power law of temperature variance $\mean{\theta^{+2}}$, which leads to the $4/3$ scaling of low-wavenumber temperature spectrum, as discussed in \S\ref{subsec_tem}. 
We use the MRF model to analyze the behavior of temperature spectrum $\phi_{\theta\theta}$. 
As shown in figure \ref{fig_MMO_theta_pdf}, the transition scales and power exponents of $\phi_{\theta\theta}$ are consistent with our analysis. 


\section{Linking MOST and MMO}
\subsection{Horizontal velocity spectrum and variance}\label{sec_u}
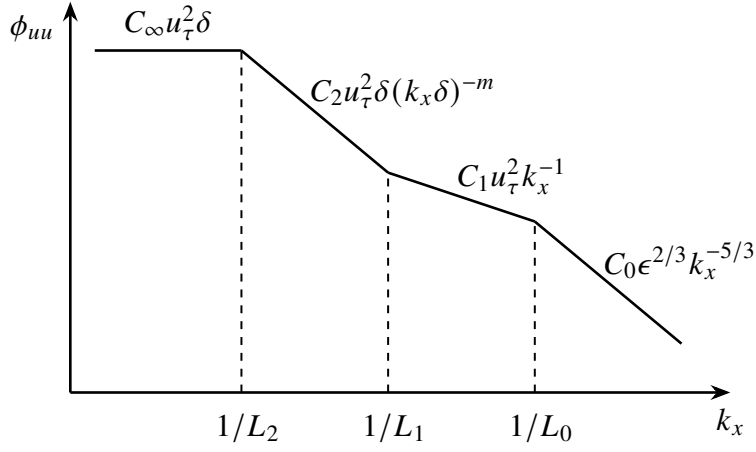
\begin{figure}[htbp]
	\centering
	\begin{tikzpicture}[xscale = 0.65, yscale = 0.65]
		\tikzstyle{every node}=[font=\normalsize]
		\draw [line width =1pt] [-Stealth] (0.5,0)--(14,0) node[yshift=0.08cm,label=-90:$k_x$]{};
		\draw [line width =1pt] [-Stealth] (0.5,0)--(0.5,8) node[xshift=0.08cm,label=190:$\phi_{uu}$]{};
            \draw[black,line width =1pt] (1,7) --(4,7);
            \node[above] at (2.5,7){$C_\infty u_\tau^2 \delta$};
            \draw[black,line width =1pt] (4,7) --(7,4.5);
            \node[above right] at (5.2,5.6){$C_2 u_\tau^2 \delta (k_x\delta)^{-m}$};
            \draw[black,line width =1pt] (7,4.5) --(10,3.5);
            \node[above right] at (8.2,3.85){$C_1 u_\tau^2 k_x^{-1}$};
            \draw[black,line width =1pt] (10,3.5) --(13,1);
            \node[above right] at (11.2,2.1){$C_0 \epsilon^{2/3}k_x^{-5/3}$};
		\draw [line width =0.75pt, dashed] (4,0) node[xshift=0.08cm,label=-90:$1/L_2$]{}--(4,7);
		\draw [line width =0.75pt, dashed] (7,0) node[xshift=0.08cm,label=-90:$1/L_1$]{}--(7,4.5);
		\draw [line width =0.75pt, dashed] (10,0) node[xshift=0.08cm,label=-90:$1/L_0$]{}--(10,3.5);
	\end{tikzpicture}
	\caption{Diagram of the multiple power law of the one-dimensional spectrum in log-log plot. }
	\label{ill_fig_1}
\end{figure}

MMO predicts three power laws for two-dimensional spectrum of horizontal velocity $u$ as listed in table \ref{table_MMO}. 
\begin{table}[htbp]
\centering
\tabcolsep=0.4cm
\renewcommand\arraystretch{1}
\begin{tabular}{cccc}
\toprule
& convective-dynamic range 
& dynamic range 
& Kolmogorov range \\
\midrule
$u$ & $-5/3$ & $-1$ & $-5/3$\\
$w$ & $1/3$ & $1$ & $-5/3$\\
$\theta$ & $-1/3$ & $-1$ & $-5/3$\\
\bottomrule
\end{tabular}
\caption{Power exponents of two-dimensional spectra predicted by MMO. The convective-dynamic range is $O(1/\delta)<k_x<O(-1/L)$, and the dynamic range is $O(-1/L)<k_x<O(1/z)$, where $k_x$ is streamwise wavenumber. }
\label{table_MMO}
\end{table}
For one-dimensional spectrum $\phi_{uu}$ we preserve the three power laws and link them with a plateau at lower wavenumber because one-dimensional spectrum peaks at zero wavenumber, as shown in figure \ref{ill_fig_1}. 
The one-dimensional spectrum is divided into four ranges:
\begin{enumerate}[(i.)]
    \item The high-wavenumber $k_x^{-5/3}$ scaling for $k_x>1/L_0$, which is the result of Kolmogorov's theory \cite{kolmogorov1941local}: 
    \begin{equation}
    \phi_{uu}=C_0 \epsilon^{2/3} k_x^{-5/3},
    \end{equation}
    where $C_0\approx 0.49$ is a universal constant \cite{pope2000turbulent}, $\epsilon$ is the turbulent energy dissipation rate, and $k_x$ is streamwise wavenumber. 
    \item The mid-wavenumber $k_x^{-1}$ scaling for $1/L_1<k_x<1/L_0$:
    \begin{equation}
        \phi_{uu}=C_1 u_\tau^2 k_x^{-1},
    \end{equation}
    which can be interpreted by Townsend's attached eddy model \cite{townsend1980structure} or asymptotic matching \cite{Nickels2005}. 
    \item The low-wavenumber $k_x^{-m}$ scaling for $1/L_2<k_x<1/L_1$:
    \begin{equation}
        \phi_{uu}=C_2 u_\tau^2 \delta (k_x\delta)^{-m}
    \end{equation}
    where $m>0$.
    \item A plateau for $k_x<1/L_2$:
    \begin{equation}
        \phi_{uu}=C_\infty u_\tau^2 \delta. 
    \end{equation}
\end{enumerate}

In boundary-layer turbulence the intermediate wavenumber $1/L_0$ between $k_x^{-5/3}$ and $k_x^{-1}$ scalings is estimated as $O(1/z)$ \cite{davidson2009simple, xie2021third}, which is the characteristic scale of attached eddies. 
In the absence of $k_x^{-m}$ scaling at low wavenumber, the lower end of $k_x^{-1}$ scaling corresponds to $O(1/\delta)$, and integrating over the wavenumber yields
\begin{equation}\label{townsend}
    \mean{u^{+2}}=\int_{0}^{\infty} \frac{\phi_{uu}}{u_\tau^2} \, \dif k_x
    \approx \int_{0}^{1/\delta} C_\infty \delta \, \dif k_x 
    + \int_{1/\delta}^{1/z} C_1 k_x^{-1} \, \dif k_x
    = C_\infty+C_1 \ln \left(\frac{\delta}{z}\right), 
\end{equation}
where $\mean{\cdot}$ denotes an average and the superscript $+$ denotes the wall-scaling normalization, and here the velocity is normalized by friction velocity $u_\tau$.  
For boundary-layer turbulence with unbalanced production and dissipation or above non-smooth walls \cite{davidson2014universal,pan2016scaling}, $L_0$ corresponds to $O(L_\epsilon)$, where $L_\epsilon=u_\tau^3/\epsilon$, because energy dissipation rate controls turbulence structure \cite{davidson2014universal,Tang2023Similarity}. 
It is further found that $L_\epsilon$ scaling is superior to $z$ scaling for second-order structure functions in convective ASL \cite{chamecki2017scaling}. 
Thus, following the process of obtaining \eqref{townsend}, a more general expression for $\mean{u^{+2}}$ is
\begin{equation}\label{townsend_revise}
    \mean{u^{+2}}=A_1+B_1 \ln \left(\frac{\delta}{L_\epsilon}\right),
\end{equation}
where $A_1$ and $B_1$ are constants related to the coefficients of spectrum. 

For strongly convective ASL, MOST presents the limiting behavior of $\mean{u^{+2}}$ corresponding to the free convection case:
\begin{equation}
    \mean{u^{+2}}\sim \frac{u_f^2}{u_\tau^2} \sim \left(-\frac{z}{L}\right)^{2/3},
\end{equation}
where $u_f=\left(\beta \mean{w\theta}_s z\right)^{1/3}$ is the local free-convection velocity, $L=-u_\tau^3 \mean{\theta}_s/(\kappa g \mean{w\theta}_s)$ with the subscript $s$ denoting the quantities calculated at the surface. 
To consistent with this limiting behavior, empirical expression for $\mean{u^{+2}}$ is proposed as
\begin{equation}\label{MOST_u}
    \mean{u^{+2}}=\left(c_1-c_2 z/L\right)^{2/3},
\end{equation}
where $c_1$ and $c_2$ are constants determined empirically. 
However, \eqref{MOST_u} does not conform well to the ASL data and, for the neutral stratification, \eqref{MOST_u} cannot be recovered to \eqref{townsend}. 
Considering the effect of $\delta$ on horizontal velocity \cite{panofsky1977characteristics, mcnaughton2006kinetic} and the significant influence of very-large-scale motions in ASL \cite{wang2016very, puccioni2023identification}, we suggest 
\begin{equation}\label{MOST_u_revise}
    \mean{u^{+2}}\sim \frac{u_f^{*2}}{u_\tau^2} \sim \left(-\frac{\delta}{L}\right)^{2/3},
\end{equation}
where $u_f^*=\left(\beta \mean{w\theta}_s \delta\right)^{1/3}$ is the free convection velocity. 
The $2/3$ power law of $-\delta/L$ is consistent with some previous research \cite{panofsky1977characteristics, Peltier1996Peltier, wilson2008monin}. 

Combining \eqref{townsend_revise} and \eqref{MOST_u_revise} we get
\begin{equation}\label{u2_revise}
    \mean{u^{+2}}=A_u^* \left(-\frac{\delta}{L}\right)^{2/3} 
    +B_u^* \ln \left(-\frac{L}{L_\epsilon}\right)+C_u^*, 
\end{equation}
where $A_u^*, B_u^*$, and $C_u^*$ are constant to be determined, and the scale $\delta$ in logarithmic part is substituted by $-L$ as we consider weak convective case $\delta>-L>z$ and thus $-L$ may be a more suitable scale. 
We will demonstrate this thereinafter. 

\eqref{u2_revise} recovers its corresponding expression in the limiting case. 
For the neutral case, $-L$ tend to infinity and should not occurs in the expression. 
Therefore, when $-L$ is greater than $\delta$, $-L$ is replaced by $\delta$ and \eqref{u2_revise} recovers to \eqref{townsend_revise}. 
With assumed local energy balance we can write $L_\epsilon$ as
\begin{equation}\label{L_ep1}
    L_\epsilon=\frac{u_\tau^3}{\epsilon}
    \approx \frac{u_\tau^3}{P+B},
\end{equation}
where $P$ is the shear production and $B$ is the buoyancy production. 
For the strong convective case with $B\approx \epsilon$ and $B=\beta \mean{w\theta}_s=u_\tau^3/(-\kappa L)$, we get $L_\epsilon\approx -\kappa L$. 
Thus \eqref{u2_revise} recovers to \eqref{MOST_u_revise}. 

\eqref{u2_revise} can also be obtained from the integration of $\phi_{uu}$. 
For with convective ASL with $z<-L<\delta$, MMO proposes that there emerges a new scaling at low wavenumber, i.e., the $k_x^{-m}$ scaling. 
Using a similar approach adopted in \citet{vassilicos2015streamwise}, we can calculate $\mean{u^2}$ from $\phi_{uu}$. 
Firstly, we match the leading order of $\phi_{uu}$ to determine coefficients $C_0, C_1$, and $C_\infty$:  
\begin{subequations}\label{coef_spectrum}
    \begin{align}
        &\text{matching at } 1/L_0: \quad 
        C_1=C_0 u_\tau^{-2} \epsilon^{2/3} L_0^{2/3}= C_0  \left(\frac{L_0}{L_\epsilon}\right)^{2/3}, \\
        &\text{matching at } 1/L_1: \quad 
        C_2=C_1 \left(\frac{L_1}{\delta}\right)^{1-m}
        =C_0 \left(\frac{L_0}{L_\epsilon}\right)^{2/3} \left(\frac{L_1}{\delta}\right)^{1-m}, \\
        &\text{matching at } 1/L_2: \quad
        C_\infty=C_2 \left(\frac{L_2}{\delta}\right)^{m}
        =C_0 \left(\frac{L_0}{L_\epsilon}\right)^{2/3} \left(\frac{L_1}{\delta}\right)^{1-m}\left(\frac{L_2}{\delta}\right)^{m}.  
    \end{align}
\end{subequations}
Then integrating $\phi_{uu}$ over $k$ yields
\begin{align}\label{u2_MMO}
    \mean{u^{+2}}\approx&\int_{0}^{1/L_2} C_\infty \delta\, \dif k_x
    +\int_{1/L_2}^{1/L_1} C_2 \delta (k_x\delta)^{-m}\, \dif k_x
    +\int_{1/L_1}^{1/L_0} C_1 k_x^{-1}\, \dif k_x
    +\int_{1/L_0}^{\infty} C_0 L_\epsilon^{-2/3}k_x^{-5/3}\, \dif k_x \notag\\
    =& C_0 \left(\frac{L_0}{L_\epsilon}\right)^{2/3}
    \left[\frac{3}{2}+\frac{1}{1-m}
    -\frac{m}{1-m}\left(\frac{L_1}{L_2}\right)^{1-m}
    +\ln \left(\frac{L_1}{L_0}\right)
    \right]. 
\end{align} 
The small scale $L_0$ should be $O(L_\epsilon)$ \cite{davidson2014universal,pan2016scaling,chamecki2017scaling}. 
According to MMO the mid scale $L_1$ is $O(-L)$ and the large scale $L_2$ reflects the outer motion in atmospheric boundary layer and corresponds to $O(\delta)$.  
The value of $m$ is $5/3$, which is consistent with the scaling in freely convective atmospheric boundary layer \cite{Kader1989, yaglom1994fluctuation}. 
Thus \eqref{u2_MMO} becomes
\begin{equation}\label{u2_MMO_2}
    \mean{u^{+2}}=C_0 
    \left[\frac{5}{2}\left(\frac{\delta}{-L}\right)^{2/3}+\ln \left(\frac{-L}{L_\epsilon}\right)\right]. 
\end{equation}
For the case when the scales and $m$ deviate from their estimates, \eqref{u2_MMO_2} can be appropriately relaxed to \eqref{u2_revise}, which is consistent with our previous argument. 

Note that \eqref{u2_revise} can be writen as
\begin{equation}
    \mean{u^{+2}}= \left(\frac{z}{\delta}\right)^{-2/3}
    \left[A_u^* \left(-\frac{z}{L}\right)^{2/3}+B_u^* \left(\frac{z}{\delta}\right)^{2/3}\ln \left(-\frac{L}{L_\epsilon}\right)+C_u^*\left(\frac{z}{\delta}\right)^{2/3}\right], 
\end{equation}
whose form has similarities to the theory proposed by \citet{stiperski2023generalizing}, i.e., the $2/3$ power of $-z/L$ multiplied by the anisotropic effect quantized by $(z/\delta)^{2/3}$. 
But our expression \eqref{u2_revise} has clearer interpretability.

\subsection{Vertical velocity spectrum and variance}
In convective ASL, the field measurements of vertical velocity variances are in good agreement with MOST:
\begin{equation}\label{MOST_w}
    \mean{w^{+2}}\sim \frac{u_f^2}{u_\tau^2} \sim \left(-\frac{z}{L}\right)^{2/3}. 
\end{equation}
However, the one-dimensional MMO spectrum of $w$ remains to be investigated, since the one-dimensional spectrum can only have negative scaling, which is inconsistent with the original MMO theory. 
\begin{table}[htbp]
\centering
\tabcolsep=0.4cm
\renewcommand\arraystretch{1}
\begin{tabular}{cccc}
\toprule
$\phi_{ww}$ & low wavenumber 
& mid wavenumber 
& high wavenumber \\
\midrule
two-dimensional MMO & $1/3$ & $1$ & $-5/3$\\
one-dimensional MMO & $-m$ & $-1$ & $-5/3$\\
\bottomrule
\end{tabular}
\caption{Two- and one-dimensional MMO scaling for $\phi_{ww}$. }
\label{table_MMO_w}
\end{table}
As listed in table \ref{table_MMO_w}, the low- and mid-wavenumber scalings need to be modified. 
At mid wavenumber, the attached eddy model accounts for the $k_x^{-1}$ scaling, which is confirmed to exist in the $\phi_{ww}$ of the measured ASL data \cite{katul1995low,Mcnaughton2007ScalingPO}. 
At low wavenumber, the scaling of $\phi_{ww}$ lacks corresponding theory, but we will show that this scaling is not important for $\mean{w^{+2}}$, at least to the leading order. 

Due to the anisotropic effect in ASL, the characteristic scale of $\phi_{ww}$ may have differences from those of $\phi_{uu}$. 
The scales $L_0$ and $L_1$ correspond to $O(0.1 z)$ and $O(z)$ respectively \cite{Mcnaughton2007ScalingPO,yang2017structure,yang2018scaling}. 
The characteristic scales of $\phi_{uu}$ are compressed compared to $\phi_{ww}$, which could be due to the fact that $w$ is controlled by local eddies, while $u$ is also affected by larger eddies. 
The successful application of \eqref{MOST_w} implies that large-scale motion has no significant effect on $w$.
Therefore, a suitable candidate for $L_2$ is $-L$, which is an intermediate scale between $z$ and $\delta$. 
According to the same approach that yielded \eqref{u2_MMO}, we can obtain the vertical velocity variance as
\begin{equation}\label{w2_MMO}
    \mean{w^{+2}}\approx
    C'_0 \left(\frac{L_0}{L_\epsilon}\right)^{2/3}
    \left[\frac{3}{2}+\frac{1}{1-m}
    -\frac{m}{1-m}\left(\frac{L_1}{L_2}\right)^{1-m}
    +\ln \left(\frac{L_1}{L_0}\right)
    \right], 
\end{equation}
where $C'_0$ is the coefficient of $k_x^{-5/3}$ scaling of $\phi_{ww}$. 
Using the approximations for production
\begin{equation}
    P\approx \frac{u_\tau^3}{\kappa z}, 
    \qquad
    B=\beta \mean{w\theta}_s=\frac{u_\tau^3}{-\kappa L},
\end{equation}
we can rewrite \eqref{L_ep1} as
\begin{equation}\label{L_epsilon_relation}
    \frac{1}{L_\epsilon} \approx \frac{1}{\kappa} \left(\frac{1}{z}+\frac{1}{-L}\right).
\end{equation}
Then we get
\begin{equation}\label{w_L0_Lep}
    \left(\frac{L_0}{L_\epsilon}\right)^{2/3}
    \sim
    \left(1-\frac{z}{L}\right)^{2/3}.
\end{equation}
To be consistent with \eqref{MOST_w} for strong convective case, we need $-\frac{m}{1-m}\left(\frac{L_1}{L_2}\right)^{1-m}$ in (\ref{w2_MMO}) to be subdominant, thus, we have a bound for the scaling exponent: $m<1$.
In Table I of the main text, the scaling exponent $m=1/2$ is obtained by the MRF model from the ASL data, which is consistent with the analytically obtained bound.
However, the specific value of $m=1/2$ remains an open question.

\subsection{Temperature spectrum and variance}\label{subsec_tem}
The temperature variance $\mean{\theta^{+2}}$ in convective ASL is described by MOST as:
\begin{equation}\label{MOST_theta}
    \mean{\theta^{+2}}\sim
    \frac{\theta_f^2}{\theta_\tau^2}\sim
    \left(-\frac{z}{L}\right)^{-2/3},
\end{equation}
where $\theta_f=\mean{w\theta}_s/u_f$ is the local free-convection temperature and $\theta_\tau=\mean{w\theta}_s/u_\tau$ is the friction temperature. 

The Kolmogorov spectrum for temperature spectrum is 
\begin{equation}\label{theta_5_3}
    \phi_{\theta\theta}=C_0'' \epsilon^{-1/3} \epsilon_\theta k_x^{-5/3},
\end{equation}
where $C_0''\approx 0.8$ is a constant \cite{wyngaard1971budgets} and $\epsilon_\theta$ is the temperature dissipation rate. 
Similar to $L_\epsilon$, we define
\begin{equation}
    L_{\epsilon_\theta}=\frac{\theta_\tau^3}{ 
    \sqrt{\epsilon_\theta^{3}/\epsilon}},
\end{equation}
and express \eqref{theta_5_3} as
\begin{equation}
    \phi_{\theta\theta}=C_0'' \theta_\tau^2 L_{\epsilon_\theta}^{-2/3} k_x^{-5/3}.
\end{equation}
Integrating $\phi_{\theta\theta}$ yields
\begin{equation}\label{theta2_MMO}
    \mean{\theta^{+2}}\approx
    C''_0 \left(\frac{L_0}{L_{\epsilon_\theta}}\right)^{2/3}
    \left[\frac{3}{2}+\frac{1}{1-m}
    -\frac{m}{1-m}\left(\frac{L_1}{L_2}\right)^{1-m}
    +\ln \left(\frac{L_1}{L_0}\right)
    \right].
\end{equation}
Considering the physical picture that heat flux is transported by plumes \cite{Mcnaughton2007ScalingPO}, the characteristic scales of $\phi_{\theta\theta}$ are analogous to that of $\phi_{ww}$. 
And from \eqref{MOST_theta} we know $\delta$ does not encounter in the expression of $\mean{\theta^{+2}}$. 
We can expect that the scales $L_0, L_1$ and $L_2$ are on the order of $0.1z, z$, and $-L$, respectively. 

If $m>1$, then from \eqref{theta2_MMO} we know $\mean{\theta^{+2}}$ is controlled by $(L_0/L_{\epsilon_\theta})^{2/3}(L_1/L_2)^{1-m}$. 
With the estimation of $\epsilon_\theta$
\begin{equation}
    \epsilon_\theta \approx \frac{u_\tau \theta_\tau^2}{\kappa_\theta z},
\end{equation}
we get 
\begin{equation}
    L_{\epsilon_\theta}\approx
    \theta_\tau^3 \epsilon^{1/2} 
    \frac{\kappa_\theta^{3/2} z^{3/2}}
    { u_\tau^{3/2} \theta_\tau^3 }
    = \frac{\kappa_\theta^{3/2} \epsilon^{1/2} 
 z^{3/2}}
    { u_\tau^{3/2} }.
\end{equation}
Thus for strong convective case, we obtain
\begin{equation}\label{z_Ltheta}
    \left(\frac{L_0}{L_{\epsilon_\theta}}\right)^{2/3}
    \sim
    \left(\frac{ u_\tau^{3/2} }
    {\epsilon^{1/2} z^{1/2}}\right)^{2/3}
    \sim
    \left(\frac{z}{ L_\epsilon }\right)^{-1/3}
    \sim \left(\frac{z}{-L}\right)^{-1/3},
\end{equation}
where \eqref{w_L0_Lep} is used. 
According to the constrain of MOST (cf. \eqref{MOST_theta}), we have
\begin{equation}
    \left(\frac{L_1}{L_2}\right)^{1-m}\approx
    \left(-\frac{z}{L}\right)^{1-m}
    \sim
    \left(-\frac{z}{L}\right)^{-1/3},
\end{equation}
therefore $m=4/3$. 
The comparison of original MMO scaling and one-dimensional MMO scaling is listed in table \ref{table_MMO_theta}. 
\begin{table}[htbp]
\centering
\tabcolsep=0.4cm
\renewcommand\arraystretch{1}
\begin{tabular}{cccc}
\toprule
$\phi_{\theta\theta}$ & low wavenumber 
& mid wavenumber 
& high wavenumber \\
\midrule
two-dimensional MMO & $-1/3$ & $-1$ & $-5/3$\\
one-dimensional MMO & $-4/3$ & $-1$ & $-5/3$\\
\bottomrule
\end{tabular}
\caption{Two- and one-dimensional MMO scaling for $\phi_{\theta\theta}$. }
\label{table_MMO_theta}
\end{table}

\bibliography{bib.bib}